\documentclass[11pt,a4paper]{article}
\usepackage{jcappub}
\usepackage[T1]{fontenc}
\usepackage{lmodern}
\usepackage{newunicodechar}
\newunicodechar{ℏ}{\ensuremath{\hbar}}
\newunicodechar{​}{}
\usepackage{microtype}
\newenvironment{subfigure}{\begingroup}{\endgroup}
\usepackage{amsmath,amssymb,bm}
\usepackage{graphicx}
\usepackage{booktabs}
\usepackage{xcolor}
\usepackage{hyperref}
\usepackage{orcidlink}
\definecolor{RED}{rgb}{1,0,0}
\newcommand{\rev}[1]{\textcolor{red}{#1}}

\graphicspath{{./}}
\hypersetup{
  colorlinks=true,
  linkcolor=blue,
  citecolor=blue,
  urlcolor=blue,
  pdftitle={Scalar Perturbations and Induced Gravitational Waves from First-Order Phase Transitions in Lattice Simulations
  },
  pdfauthor={Yuan-Jie Li, Jing Liu, Zhuan Ning, Xiang-Xi Zeng, Zong-Kuan Guo}
}

\allowdisplaybreaks
\def\INTEGRATEDMANUSCRIPT{1}

\title{Scalar Perturbations and Induced Gravitational Waves from First-Order Phase Transitions in Lattice Simulations
}
\author[a,b]{Yuan-Jie Li,}
\author[a,b]{Jing Liu,}
\author[c,d,e]{Zhuan Ning,}
\author[f,g]{Xiang-Xi Zeng\,\orcidlink{0009-0003-0632-3230},}
\author[d,h,c]{and Zong-Kuan Guo}

\affiliation[a]{International Centre for Theoretical Physics Asia-Pacific,\protect\\
University of Chinese Academy of Sciences, 100190 Beijing, China}
\affiliation[b]{Taiji Laboratory for Gravitational Wave Universe,\protect\\
University of Chinese Academy of Sciences, 100049 Beijing, China}
\affiliation[c]{School of Fundamental Physics and Mathematical Sciences, Hangzhou Institute for Advanced Study,\protect\\
University of Chinese Academy of Sciences, Hangzhou 310024, China}
\affiliation[d]{Institute of Theoretical Physics, Chinese Academy of Sciences,\protect\\
P.O. Box 2735, Beijing 100190, China}
\affiliation[e]{University of Chinese Academy of Sciences, Beijing 100049, China}
\affiliation[f]{State Key Laboratory of Dark Matter Physics, Tsung-Dao Lee Institute and School of Physics and Astronomy,\protect\\
Shanghai Jiao Tong University, Shanghai 201210, China}
\affiliation[g]{Key Laboratory for Particle Astrophysics and Cosmology (MOE), and Shanghai Key Laboratory for Particle Physics and Cosmology,\protect\\
Shanghai Jiao Tong University, Shanghai 201210, China}
\affiliation[h]{School of Physical Sciences, University of Chinese Academy of Sciences,\protect\\
No.19A Yuquan Road, Beijing 100049, China}

\emailAdd{liyuanjie23@mails.ucas.ac.cn}
\emailAdd{liujing@ucas.ac.cn}
\emailAdd{ningzhuan17@mails.ucas.ac.cn}
\emailAdd{zengxiangxi@sjtu.edu.cn}
\emailAdd{guozk@itp.ac.cn}
\date{\today}

\abstract{
{Cosmological first-order phase transitions can generate curvature perturbations through inhomogeneous quantum tunneling, as studied previously on superhorizon scales.}
In this work, we for the first time conduct three-dimensional lattice simulations that incorporate scalar metric perturbations and radiation perturbations, covering a range from bubble wall scales to super-horizon scales. We obtain the precise scalar perturbation power spectrum and the probability density function  of energy density perturbations. Furthermore, we simulate the {gravitational-wave energy spectra} generated by each source during the {first-order phase transition}, including the scalar field itself, scalar metric perturbations, and the energy density and velocity perturbations of radiation. For {gravitational waves}, the contribution from other sources can exceed $1/3$ of that from the scalar field at superhorizon scales. Additionally, we compare the effects of different values of the transition strength and rate on the results. This paper provides more accurate numerical results for research aimed at detecting or constraining {first-order phase transitions} via {gravitational waves} and curvature perturbations.
}

\begin{document}
\maketitle

\ifdefined\INTEGRATEDMANUSCRIPT
\else
\documentclass[11pt,a4paper]{article}


\usepackage[T1]{fontenc}
\usepackage{lmodern}
\usepackage{microtype}
\usepackage[a4paper,margin=25mm]{geometry}

\usepackage{amsmath}
\usepackage{amssymb}
\usepackage{bm}
\usepackage{graphicx}
\usepackage{booktabs}
\usepackage{cite}

\usepackage{xcolor}
\usepackage{hyperref}
\definecolor{RED}{rgb}{1,0,0}
\newcommand{\rev}[1]{{#1}}

\hypersetup{
  colorlinks = true,
  linkcolor  = blue,
  citecolor  = blue,
  urlcolor   = blue,
  pdftitle   = {Metric and Radiation Perturbations and Induced Gravitational Waves from First-Order Phase Transitions},
  pdfauthor  = {}
}

\setlength{\emergencystretch}{2em}
\allowdisplaybreaks

\title{\bfseries
  \rev{Metric and Radiation Perturbations and Induced Gravitational Waves
  from First-Order Phase Transitions in Three-Dimensional Lattice Simulations}
}
\author{}
\date{}

\begin{document}

\maketitle

\fi

\section{Introduction}
\label{sec:introduction}

Cosmological first-order phase transitions (FOPTs) can connect particle physics beyond the Standard Model to observable relics in the early Universe \cite{Athron:2023xlk,Caprini:2019egz,Liu:2021jyc,Wang:2024slx,Tian:2024ysd}. Their dynamics proceeds through bubble nucleation, expansion, and collision \cite{Ai:2021kak,Ai:2023see,Wang:2025eee,Lu:2022paj,Cai:2018teh}. Three-dimensional simulations have extensively studied the {resulting gravitational-wave (GW) production} \cite{Li:2023yzq,Di:2020kbw,Zhao:2022cnn,Bian:2025twi,Bian:2026xdm,Zou:2026wzi}, including long-lived acoustic sources in the plasma \cite{Hindmarsh:2013xza,Hindmarsh:2015qta,Wang:2026jnh} and vacuum bubble collisions \cite{Cutting:2018tjt}. Modern lattice frameworks such as \textsc{CosmoLattice} can now follow nonlinear field dynamics and tensor perturbations in an expanding Universe \cite{Figueroa:2020rrl,Figueroa:2021yhd}.
{These capabilities provide a foundation for extending lattice calculations to include scalar metric and radiation perturbations and {the gravitational waves they induce (IGWs)}.}

Scalar perturbations generated during a FOPT provide a complementary probe of its inhomogeneous dynamics. The stochastic nucleation and subsequent evolution of bubbles induce spatial variations in the local energy density \cite{Cai:2024nln}. These perturbations are further communicated to the radiation fluid through gravity, generating coupled fluctuations in the scalar field, spacetime, and plasma. Such scalar perturbations can leave observable imprints in two distinct ways: directly through the resulting density and curvature perturbations, which may have implications for {primordial black hole (PBH) production} \cite{Turner:1992tz,Liu:2022lvz,Elor:2023xbz,Liu:2021svg,Kawana:2022olo,Lewicki:2023ioy,Gouttenoire:2023naa,Kanemura:2024pae,Ai:2026zrs,Ai:2024cka,Wang:2025hwc,Ning:2026nfs}, and indirectly through {IGWs} sourced at second order \cite{Ananda:2006af,Baumann:2007zm,Kohri:2018awv,Espinosa:2018eve,Cai:2019amo}. A self-consistent treatment of these effects therefore requires evolving the scalar field, radiation fluid, and scalar metric perturbations simultaneously throughout the FOPT.

Curvature perturbations generated by {FOPTs} have been studied under assumptions such as superhorizon evolution, rare delayed regions, and spherical symmetry \cite{Franciolini:2025ztf,Wang:2026zvz,Giombi:2023jqq,Jinno:2024nwb,Cai:2024nln}, while post-inflationary lattice studies have further investigated the density perturbations generated by phase-transition dynamics in a homogeneous radiation background \cite{Zou:2025sow}. Scalar metric perturbations have also been incorporated in three-dimensional lattice simulations of nonlinear preheating \cite{Huang:2011gf}, while a FOPT occurring during inflation has been simulated with the accompanying scalar metric and curvature perturbations \cite{An:2023jxf}. Separately, lattice calculations of {IGWs} have been developed in the context of nonlinear structure formation, non-Gaussian or isocurvature primordial perturbations, and nonlinear inflationary dynamics \cite{Fernandez:2023ddy,Zeng:2025cer,Zeng:2025tno,Caravano:2026hca}. In this work, we simulate a FOPT during the radiation dominated era while consistently evolving the scalar field, radiation fluid, and scalar metric perturbations. To our knowledge, this is the first three-dimensional multi-bubble simulation of a post-inflationary FOPT that, within linear metric perturbation theory, simultaneously determines both scalar metric potentials together with the radiation density and velocity perturbations. It is also the first such simulation to obtain the {IGW} spectrum directly from the local, dynamically generated source.

We perform three-dimensional simulations in the Newtonian gauge, where scalar metric perturbations are described by the Newtonian potential $\Phi$ and the spatial-curvature potential $\Psi$. We assume no direct coupling between the scalar field and radiation, where radiation density and velocity perturbations are dynamically generated through their gravitational coupling to $\Phi$ and $\Psi$. {This setup allows us to isolate the gravitationally induced radiation perturbations and their GW contribution from the conventional sound-wave contribution driven by direct scalar-field--fluid interactions.} We characterize the evolution of the curvature-perturbation power spectrum, the statistical distribution of the horizon-scale density perturbations, and the final GW energy spectra sourced by the scalar field, radiation, and metric perturbations. We further examine how these observables depend on the two basic phase-transition parameters, the transition strength {$\alpha$} and inverse duration {$\beta$}. We find that the sources other than the scalar field become increasingly important around the horizon scale, contributing more than $1/3$ of the scalar-field contribution in part of the GW energy spectrum. Their inclusion also modifies the infrared behavior of the total GW energy spectrum, producing a frequency-dependent spectral index rather than a single standard power law. These predictions provide multiple avenues for observational tests. The {GW} signal can be probed across complementary frequency bands, while the accurately predicted curvature perturbations may be constrained independently through cosmic microwave background spectral distortions and the formation of compact dark-matter structures.

This paper is organized as follows. {Section~\ref{sec:setup} introduces the phase-transition model and
the equations governing the scalar field, radiation, and metric
perturbations.
With this framework in place, we describe the numerical setup and
examine the scalar perturbations, density statistics, and GW spectra
in Sec.~\ref{sec:metric-density-results}.
We summarize our findings and discuss future directions in
Sec.~\ref{sec:discussion-conclusions}.
Details of the bubble-nucleation prescription, perturbation evolution,
GW computation, and statistical estimators are collected in
Appendices~\ref{app:exponential-nucleation-module}--\ref{app:power-spectrum-pdf-module}.} For convenience, we use $c=\text{ℏ}=1$ throughout this paper.

\ifdefined\INTEGRATEDMANUSCRIPT
\else
\bibliographystyle{unsrt}
\bibliography{paper_introduction_references}

\end{document}
\fi

\ifdefined\INTEGRATEDMANUSCRIPT
\else
\documentclass[11pt,a4paper]{article}

\usepackage[T1]{fontenc}
\usepackage{lmodern}
\usepackage{microtype}
\usepackage[a4paper,margin=25mm]{geometry}
\usepackage{amsmath}
\usepackage{amssymb}
\usepackage{bm}
\usepackage{cite}
\usepackage{xcolor}
\usepackage{hyperref}
\definecolor{RED}{rgb}{1,0,0}
\newcommand{\rev}[1]{{#1}}

\hypersetup{
  colorlinks = true,
  linkcolor  = blue,
  citecolor  = blue,
  urlcolor   = blue
}

\setlength{\emergencystretch}{2em}
\allowdisplaybreaks

\begin{document}
\fi

\section{The model and the equation of motion}
\label{sec:setup}

\subsection{The FOPT model and the bubble nucleation}
\label{sec:setup-transition}

We consider a real scalar field $\phi$ in an expanding Universe, where the action reads
\begin{equation}
  S=\int d^4x\sqrt{-g}\left[
  \frac{1}{16\pi G}\mathcal R
  -\frac12g^{\mu\nu}\partial_\mu\phi\partial_\nu\phi
  -V(\phi)\right]+S_r,
  \label{eq:setup-action}
\end{equation}
where $g_{\mu\nu}$ is the spacetime metric with its determinant $g$, and $\mathcal R$
is the Ricci scalar, and $S_r$ refers to the background radiation component.
The corresponding Klein--Gordon equation of $\phi$ reads
\begin{equation}
  \phi''+2\mathcal H\phi'-\nabla^2\phi+a^2 V_{,\phi}=0,
  \label{eq:setup-klein-gordon}
\end{equation}
where {$a$ is the scalar factor}, a prime denotes a derivative respective to the conformal time $\tau$, and $\mathcal H\equiv a'/a$ is {the conformal Hubble parameter}. 
{The background expansion is determined self-consistently by the
scalar field and radiation through the Friedmann equation,}
\begin{align}
	\mathcal H^2
	&=\frac{8\pi G}{3}a^2
	\left(\bar\rho_\phi+\bar\rho_r\right),
	\nonumber\\
	\bar\rho_\phi
	&=\left\langle
	\frac{\phi'^2+|\nabla\phi|^2}{2a^2}+V(\phi)
	\right\rangle.
	\label{eq:setup-friedmann}
\end{align}
{Spatial derivatives are taken with respect to comoving coordinates,
while angle brackets and overbars denote spatial averages over the simulation volume.
The quantity $\bar\rho_\phi$ is the mean
scalar-field energy density, evaluated from the evolving field configuration. The scalar-field energy includes kinetic, gradient, and potential contributions.}
The energy density of radiation scales as $\bar\rho_r {\propto} a^{-4}$.

We apply the following potential to induce the FOPT
\begin{equation}
  V(\phi)
  =
  \frac{\lambda_\phi}{4}\phi^4
  -\frac{\beta_\phi}{3}\phi^3
  +\frac{\mu_\phi^2}{2}\phi^2
  +V_0,
  \label{eq:setup-potential}
\end{equation}
Here $\lambda_\phi$, $\beta_\phi$, and $\mu_\phi$ are the coefficients of the quartic, cubic, and
quadratic terms, respectively. The constant $V_0$ is chosen so that the potential energy of the true-vacuum vanishes. Hence the
released vacuum energy during the FOPT is $\rho_{\rm vac}=V(\phi_{\rm false})-V(\phi_{\rm true})$, where $\phi_{\rm false}$ and $\phi_{\rm true}$ denote the vacuum expectation value of the false and true vacua. 

The field $\phi$ settles in the false vacuum at the initial time. 
Vacuum tunneling initiates the nucleation of true-vacuum bubbles, with the released latent energy concentrated in the bubble walls as gradient and kinetic energy. The expansion and collision of bubbles generate the density inhomogeneity and anisotropic stress that source
metric perturbations studied below. 
After bubble collisions, the resulting
relativistic scalar excitations can behave approximately as a radiation-like
component at the coarse-grained level, but they are not injected directly
into the separately evolved radiation fluid in our case, i.e., the background radiation couple to $\phi$ only through the gravitational interaction.  This separates the gravitationally generated perturbations in radiation studied here from the acoustic fluid waves produced by direct
wall--plasma energy and momentum transfer in scalar--fluid simulations
\cite{Hindmarsh:2013xza,Hindmarsh:2015qta}.  We define the transition strength
at the reference nucleation time $\tau_n$ as
\begin{equation}
  \alpha
  \equiv
  \frac{\rho_{\rm vac}}{\bar\rho_r(\tau_n)}\,,
  \label{eq:setup-alpha}
\end{equation}
which denotes the ratio between the energy densities of vacuum energy and the radiation.

Bubble nucleation is treated as an inhomogeneous Poisson process \cite{Turner:1992tz}, { with a nucleation rate that grows exponentially
in physical time}. The tunneling rate per physical volume and physical time is
\begin{equation}
  \Gamma_{\rm phys}(t)
  =H_n^4\exp\!\left[\beta_{\rm nuc}(t-t_n)\right],
  \qquad
  \beta_{\rm nuc}
  =\left(\frac{\beta}{H}\right)_n H_n,
  \label{eq:setup-nucleation-rate}
\end{equation}
where $dt=a\,d\tau$, $H_n=a'(\tau_n)/a^2(\tau_n)$, and
$t_n=\int^{\tau_n}a\,d\tau$. 

The two basic parameters $\alpha$ and $(\beta/H)_n$ play complementary
	roles in the FOPT, with the former quantifying the transition strength through the released latent heat and the latter measuring the inverse time duration of bubble nucleation.

The critical bubble of the chosen transition model is determined by the
bounce solution that extremizes the Euclidean action.  We represent its radial
profile as
\begin{equation}
  \delta\phi_{\rm bub}(r_{\rm phys})
  =\frac{\eta}{2}
  \left[1-\tanh\!\left(\frac{r_{\rm phys}-R_c}{\ell_0}\right)\right],
  \label{eq:setup-bubble-profile}
\end{equation}
where $\eta\equiv |\phi_{\rm true}-\phi_{\rm false}|${,} the critical radius $R_c$ and wall thickness $\ell_0$ are fixed by {the} bounce solution, and $r_{\rm phys}$ is the radius in the physical coordinate.

\subsection{Scalar perturbations of the metric and the radiation energy density}
\label{sec:setup-scalar-perturbations}

We describe scalar perturbations in conformal Newtonian gauge,
\begin{equation}
  ds^2=a^2(\tau)\left[-(1+2\Phi)d\tau^2+(1-2\Psi)\delta_{ij}dx^i dx^j\right],
  \label{eq:setup-newtonian-gauge}
\end{equation}
Here $\Phi$ is the Newtonian potential and $\Psi$ is the
spatial-curvature potential \cite{Ma:1995ey,Baumann:2007zm}.  
Due to the non-negligible anisotropic stress arising from nonlinear processes such as bubble collisions in {FOPT} simulations, we do not impose $\Phi=\Psi$. Instead, we treat $\Phi$ and $\Psi$ as independent variables.

The radiation density contrast is defined as
$\delta_r\equiv\delta\rho_r/\bar\rho_r$, and the radiation velocity divergence
is then $\theta_r\equiv\partial_i v_r^i$. We define the scalar velocity potential by
$v_{r,i}=-\partial_i v_r$, so that
$\theta_r=-\nabla^2v_r$. 
{We denote Fourier modes by $X_{\bm{k}}\equiv\mathcal{F}[X](\bm{k})$,
where $\mathcal{F}$ is the spatial Fourier transform with kernel
$e^{-i\bm{k}\cdot\bm{x}}$, $\bm{k}$ is the comoving wavevector,
and $k=|\bm{k}|$. Thus, $\nabla^2\to-k^2$ and
$v_{r,\bm{k}}=\theta_{r,\bm{k}}/k^2$ for $k\ne0$.}
Under the assumption $p_r=\rho_r/3$, adiabatic sound speed
$c_s^2=1/3$, and vanishing shear, we linearize the energy-momentum conservation equation $\nabla_\mu T_r^{\mu\nu}=0$ as \cite{Ma:1995ey}
\begin{equation}
  \delta'_{r,\bm{k}}
  =-\frac43\theta_{r,\bm{k}}+4\Psi'_{\bm{k}},
  \qquad
  \theta'_{r,\bm{k}}
  =k^2\left(\frac14\delta_{r,\bm{k}}+\Phi_{\bm{k}}\right).
  \label{eq:setup-radiation-equations}
\end{equation}
The first equation of Eq.~\eqref{eq:setup-radiation-equations} is the perturbed continuity equation, where radiation is
redistributed by the velocity divergence, while the spatial curvature
affects the local volume element through the $4\Psi'$ term. The second one is the
Euler equation, where the pressure gradient, represented by $\delta_r/4$, and the
Newtonian potential $\Phi$ accelerate the fluid. Radiation therefore behaves as a
gravitationally driven acoustic component rather than as an independently
prescribed density profile.
Note that, although we have included scalar metric perturbations in this subsection, we neglect their backreaction on the dynamics of the scalar field $\phi$, described by Eq.~\eqref{eq:setup-klein-gordon}, which remains predominantly governed by its strong self-interaction and the background expansion of the universe. 


The evolution of scalar metric perturbations is in turn determined by the energy-momentum tensor of $\phi$ and radiation.
For every nonzero Fourier mode we then define
\begin{align}
  B_{\bm{k}}
  &\equiv4\pi G a^2
  \left(\delta\rho_{\phi,\bm{k}}
  +\bar\rho_r\delta_{r,\bm{k}}\right),
  \nonumber\\
  M_{\bm{k}}
  &\equiv4\pi G\left[
  q_{\phi,\bm{k}}
  +\frac{4a^2\bar\rho_r}{3k^2}\theta_{r,\bm{k}}
  \right],
  \nonumber\\
  A_{\bm{k}}
  &\equiv8\pi G a^2\Pi^S_{\phi,\bm{k}}.
  \label{eq:setup-source-combinations}
\end{align}
where $\delta\rho_{\phi,\bm{k}}$ denotes density perturbations of $\phi$, the longitudinal momentum potential $q_{\phi,\bm{k}}$ is defined by $\delta T^0{}_{i,\phi}=-a^{-2}\partial_iq_\phi$, and the scalar anisotropic stress $\Pi^S_{\phi,\bm{k}}$
\begin{equation}
	\Pi^S_{\phi,\bm{k}}
	\equiv\frac32 Q_i{}^j\Sigma^i{}_{j,\phi}(\bm{k}),
	\qquad
	Q_i{}^j\equiv\hat k_i\hat k^j-\frac13\delta_i{}^j,
	\label{eq:setup-anisotropic-stress-convention}
\end{equation}
where $\Sigma^i{}_{j,\phi}\equiv\delta T^i{}_{j,\phi}-\delta^i{}_j\delta T^k{}_{k,\phi}/3$ is the traceless scalar-field spatial stress. We also use the Fourier convention $\nabla^2\rightarrow-k^2$.

Eq.~\eqref{eq:setup-source-combinations} includes three physical effects. $B_{\bm{k}}$ represents the total
energy density perturbations, $M_{\bm{k}}$ is the longitudinal momentum flow of the scalar field and radiation, and $A_{\bm{k}}$ is the scalar projection of the anisotropic stress of $\phi$. 
Correspondingly, the $00$, scalar $0i$, and traceless spatial components of the linearized
Einstein equations, $\delta G^\mu{}_{\nu}=8\pi G\delta T^\mu{}_{\nu}$, then read
\cite{Ma:1995ey}
\begin{align}
  -k^2\Psi_{\bm{k}}
  -3\mathcal H\left(\Psi'_{\bm{k}}+\mathcal H\Phi_{\bm{k}}\right)
  &=B_{\bm{k}},
  \nonumber\\
  \Psi'_{\bm{k}}+\mathcal H\Phi_{\bm{k}}
  &=M_{\bm{k}},
  \nonumber\\
  k^2\left(\Phi_{\bm{k}}-\Psi_{\bm{k}}\right)
  &=A_{\bm{k}}.
  \label{eq:setup-einstein-constraints}
\end{align}
Substituting the momentum constraint into the $00$ equation and then using
the traceless constraint gives the algebraic projection
\begin{align}
  \Psi_{\bm{k}}
  &=-\frac{B_{\bm{k}}+3\mathcal H M_{\bm{k}}}{k^2},
  \nonumber\\
  \Phi_{\bm{k}}
  &=\Psi_{\bm{k}}+\frac{A_{\bm{k}}}{k^2},
  \nonumber\\
  \Psi'_{\bm{k}}
  &=M_{\bm{k}}-\mathcal H\Phi_{\bm{k}}.
  \label{eq:setup-einstein-projection}
\end{align}
Equation~\eqref{eq:setup-einstein-projection} implies that the energy density and momentum perturbations jointly determine the spatial curvature $\Psi$, the momentum constraint determines its time evolution, and the
anisotropic stress leads to the difference between $\Phi$ and $\Psi$. 
Equation \eqref{eq:setup-einstein-projection} is applied throughout the nonlinear evolution in numerical simulations.
{Within our approximation scheme, the reconstructed quantities
$\Phi$, $\Psi$, and $\Psi'$ provide all the scalar metric information
required to evolve the scalar-field--radiation system and compute
the observables considered here.
We therefore do not solve the remaining scalar Einstein equations
separately.}

We note that the evolution of $\phi$ is highly nonlinear which requires lattice simulations, while perturbations of the metric and radiation remains at the linear order. As shown below, $|\Phi|\ll1$ and $|\Psi|\ll1$ throughout the resolved evolution, which supports the self-consistency of this perturbative sector. {The small amplitudes of these potentials justify neglecting their
backreaction on the scalar-field dynamics at leading order.}

For comparison with a gauge-invariant
curvature variable, we additionally construct the curvature perturbation on
uniform-density hypersurfaces,
$\zeta=-\Psi-\mathcal H\delta\rho_{\rm tot}/\bar\rho'_{\rm tot}$. In intervals
where total background energy conservation gives
$\bar\rho'_{\rm tot}=-3\mathcal H
(\bar\rho_{\rm tot}+\bar p_{\rm tot})$, this becomes \cite{Wands:2000dp}
\begin{equation}
	\zeta
	=-\Psi-\mathcal H\frac{\delta\rho_{\rm tot}}{\bar\rho'_{\rm tot}}
	=-\Psi
	+\frac{\delta\rho_\phi+\bar\rho_r\delta_r}
	{3\left(\bar\rho_\phi+\bar p_\phi+\frac43\bar\rho_r\right)}.
	\label{eq:setup-zeta}
\end{equation}
On scales much larger than the Hubble radius, $\zeta$ is conserved for an
adiabatic system with negligible gradient terms. During the transition,
however, nonadiabatic pressure and the discrete field-injection prescription
limit that interpretation. 



Our primary scalar observables are the dimensionless power spectra of the
two potentials $\Phi$ and $\Psi$. Their difference diagnoses the scalar
gravitational slip. 

For any scalar $X$, we define
\begin{align}
  \langle X_{\bm{k}}X^*_{\bm{k}'}\rangle
  &=(2\pi)^3\delta^{(3)}(\bm{k}-\bm{k}')P_X(k),
  \nonumber\\
  \Delta_X^2(k)&=\frac{k^3}{2\pi^2}P_X(k).
  \label{eq:setup-power-spectrum}
\end{align}
The factor $k^3/(2\pi^2)$ follows from integrating an isotropic three-dimensional
Fourier spectrum over angular directions, so that
$\langle(X-\langle X\rangle)^2\rangle=\int d\ln k\,\Delta_X^2(k)$.
Our lattice estimator follows the shell-averaged Fourier convention used by
\textsc{CosmoLattice} \cite{Figueroa:2020rrl,Figueroa:2021yhd}.
We use $\Delta_\Phi^2(k)$ and $\Delta_\Psi^2(k)$ as the principal spectra,
$\Delta_{\Phi-\Psi}^2(k)$ as a slip diagnostic, and $\zeta$ for
supplementary comparisons. The dimensionless spectrum is the
contribution to the variance per logarithmic wavenumber interval, {revealing which length scales dominate the perturbation.} 

To characterize the energy density perturbations, we also use the total
density contrast in the comoving gauge~\footnote{We adopt the comoving gauge here mainly to facilitate comparison with previous studies, such as Ref.~{\cite{Wang:2026zvz}}.}
\begin{align}
  \delta_{c,\bm{k}}&={1\over\bar\rho_\phi+\bar\rho_r}[\delta\rho_{\phi,\bm{k}}+\bar\rho_r\delta_{r,\bm{k}}+{3\mathcal H\over a^2}
  \left(q_{\phi,\bm{k}}+
  {4a^2\bar\rho_r\over3k^2}\theta_{r,\bm{k}}\right)\Biggr],
  \qquad \bm{k}\ne\bm{0},
  \label{eq:setup-delta-comoving}
\end{align}
where $\delta T^0{}_{i,\bm{k}}=-ik_iq_{\bm{k}}/a^2$ and $q_{\bm{k}}=q_{\phi,\bm{k}}+4a^2\bar\rho_r\theta_{r,\bm{k}}/(3k^2)$.
To characterize density perturbations on a specific scale $R$, we apply $\delta_c^{(R)} = \mathcal{F}^{-1}[W(kR)\delta_{c,\mathbf{k}}]$ with a Gaussian kernel $W(kR) = \exp(-k^2R^2/2)$, from which we then derive the corresponding {normalized probability density function (PDF)}.


\subsection{Tensor metric perturbations}
\label{sec:setup-tensor}

We study the tensor perturbations generated by the scalar-field dynamics and the perturbations induced in the surrounding radiation fluid. 

{Including tensor perturbations, we write the metric in conformal
Newtonian gauge as
\begin{equation}
  ds^2=a^2(\tau)\left\{
  -(1+2\Phi)d\tau^2+
  \left[(1-2\Psi)\delta_{ij}+h_{ij}\right]dx^i dx^j
  \right\},
  \label{eq:setup-metric-tensor}
\end{equation}
where $h_{ij}$ is transverse and traceless,
$\partial_i h_{ij}=0$ and $\delta^{ij}h_{ij}=0$.} The tensor perturbation \(h_{ij}\) obeys
\begin{equation}
	h_{ij}^{\prime\prime}
	+2\mathcal H h_{ij}^{\prime}
	-\nabla^2 h_{ij}=
	\Lambda_{ij}{}^{lm}
	\left(
	S_{lm}^{\rm s}
	+S_{lm}^{\rm g}
	+S_{lm}^{\rm r}
	\right),
	\label{eq:setup-gw}
\end{equation}
where \(\Lambda_{ij}{}^{lm}\) denotes the transverse--traceless (TT) projector. The three source terms describe the scalar-field contribution, the nonlinear geometric contribution from scalar metric perturbations, and the contribution from perturbations in the radiation fluid, respectively.

The scalar-field GW source is
\begin{equation}
	S_{ij}^{\rm s}
	=16\pi G\,(\partial_i{\phi})(\partial_j{\phi}).
	\label{eq:setup-phi-source}
\end{equation}
Only the anisotropic part of this stress contributes after TT projection, while the isotropic pressure and vacuum energy do not source tensor perturbations.

The geometric source arises from the nonlinear coupling of the first-order scalar metric perturbations{~\cite{Baumann:2007zm}}. In terms of the scalar potentials \(\Phi\) and \(\Psi\), it is
\begin{align}
	S_{ij}^{\rm g}
	={-2} \,\big[&
	2\Phi\partial_i\partial_j\Phi
	-2\Psi\partial_i\partial_j\Phi
	+4\Psi\partial_i\partial_j\Psi
	+(\partial_i\Phi)(\partial_j\Phi)\nonumber\\&
	-(\partial_i\Phi)(\partial_j\Psi)
	-(\partial_i\Psi)(\partial_j\Phi)
	+3(\partial_i\Psi)(\partial_j\Psi)
	\big].
	\label{eq:setup-geometric-source}
\end{align}

The perturbed radiation fluid provides an additional source through its quadratic velocity contribution,
\begin{equation}
	S_{ij}^{\rm r}
	=\frac{{64}\pi G}{3}a^2\bar\rho_r
	(\partial_i v_r)(\partial_j v_r),
	\qquad
	v_{r,\mathbf{k}}=\frac{\theta_{r,\mathbf{k}}}{k^2},
	\label{eq:setup-radiation-source}
\end{equation}
where \(\bar\rho_r\) is the homogeneous radiation density. This term accounts for the tensor source associated with the perturbed radiation fluid.

\ifdefined\INTEGRATEDMANUSCRIPT
\else
\bibliographystyle{unsrt}
\bibliography{paper_introduction_references}

\end{document}
\fi

\section{Simulations and results}
\label{sec:metric-density-results}

\subsection{The numerical setup}
\label{sec:numerical-setup}

To realize the evolution of $\phi$ in the lattice, we use the same field and frequency normalization of
Ref.~\cite{Li:2025zxa}. The reference scales are
\begin{equation}
	f_*\equiv\eta,
	\qquad
	\omega_*\equiv\sqrt{\lambda_\phi}\,\eta\,.
	\label{eq:setup-normalization-scales}
\end{equation}
The dimensionless quantities employed in the program (field value, time, length, and momentum) are scaled respect to $f_*$ and $\omega_*$.  We apply
\begin{equation}
	\varphi\equiv\frac{\phi}{f_*},
	\qquad
	\widehat\tau\equiv\omega_*\tau,
	\qquad
	\widehat{\bm{x}}\equiv\omega_*\bm{x},
	\qquad
	\widehat k\equiv\frac{k}{\omega_*}\,.
	\label{eq:setup-dimensionless-variables}
\end{equation}

The dimensionless potential used in the simulations is defined by
\begin{equation}
	\widehat V(\varphi)
	\equiv\frac{V(f_*\varphi)}{f_*^2\omega_*^2}\,,
	\label{eq:setup-dimensionless-potential}
\end{equation}
and the dimensionless parameters in $\widehat V(\varphi)$ are chosen as
\begin{equation}
	\left(
	\frac{\lambda_\phi f_*^2}{\omega_*^2},
	\frac{\beta_\phi f_*}{\omega_*^2},
	\frac{\mu_\phi^2}{\omega_*^2},
	\frac{V_0}{f_*^2\omega_*^2}
	\right)
	=\left(1,1.3,0.3,\frac1{30}\right),
	\label{eq:setup-dimensionless-couplings}
\end{equation}
or equivalently
\begin{equation}
	\widehat V(\varphi)
	=\frac14\varphi^4
	-\frac{1.3}{3}\varphi^3
	+\frac{0.3}{2}\varphi^2
	+\frac1{30}\,,
	\label{eq:setup-dimensionless-potential-parameters}
\end{equation}
which gives $\varphi_{\rm false}=0$ and $\varphi_{\rm true}=1$. 


We perform FOPT simulations on a $512^3$ lattice with $L\omega_*=256$, $\Delta\tau\,\omega_*=0.1$, $R_c\omega_*=10.06$, and $\ell_0\omega_*=3.05$. With these parameters, the wall thickness is much smaller than the typical bubble collision scale. For example, at insertion where $a\simeq1$, the wall thickness spans approximately $\ell_0/\Delta x\simeq6.1$ lattice intervals, with $\Delta x\,\omega_*=1/2$. Therefore, the lattice resolution is sufficient for our subsequent spectral analysis focusing on the typical scales of curvature perturbations and {GWs}, which are safely separated from the wall-scale ultraviolet structure.

The process of quantum tunneling is realized in the lattice self-consistently with Eq.~\eqref{eq:setup-nucleation-rate}, see Appendix~\ref{app:exponential-nucleation-module} for details. {In our simulations, we set the reference nucleation time to
$\tau_n\omega_*=20$.} Among the simulations, we compare the numerical results with different values of $\alpha$ and $(\beta/H)_n$.

For comparisons among simulations we denote all lengths in this
paragraph as comoving and characterize the mean separation of nucleation centers by
\begin{equation}
	R_*
	=\left(\frac{L^3}{N_{\rm bub}}\right)^{1/3}
	=n_{\rm bub}^{-1/3},
	\label{eq:setup-bubble-separation}
\end{equation}
where $L$ is the comoving box length, $N_{\rm bub}$ is the number of
accepted bubbles after nucleation has ceased, and $n_{\rm bub}=N_{\rm bub}/L^3$
is their final comoving number density.

To avoid the computational cost of performing the transverse-traceless (TT) projection at every time step, we first evolve the unprojected perturbation $u_{ij}$, which equation of motion is given by
	\begin{equation}
		u_{ij}^{\prime\prime}
		+2\mathcal H u_{ij}^{\prime}
		-\nabla^2u_{ij}
		= \frac{{1}}{g} \left( S_{ij}^{\rm s} + S_{ij}^{\rm g} + S_{ij}^{\rm r} \right),
		\label{eq:total-gw}
\end{equation}
where the gravitational normalization is carried by the dimensionless ratio $g=(f_*/M_{\rm Pl})^2$. The physical TT perturbation is then obtained via $h_{ij}=g\Lambda_{ij}{}^{lm}u_{lm}$ when estimating the spectrum, with $\Lambda_{ij}{}^{lm}$ being the Fourier-space TT projector. The GW sources used in the program, ${S}_{ij}^{\rm s}$, ${S}_{ij}^{\rm g}$, ${S}_{ij}^{\rm r}$, are defined from Eqs.~(\eqref{eq:setup-phi-source},\eqref{eq:setup-geometric-source},\eqref{eq:setup-radiation-source}) using the lattice quantities and operators.

The dimensionless GW energy spectrum is defined as 
	\begin{equation}
		\Omega_{\rm GW}(k)\equiv
		\frac{1}{\bar\rho_{\rm tot}}\frac{d\rho_{\rm GW}} {d\ln k},
	\end{equation}
namely the GW energy per logarithmic comoving-wavenumber interval divided by the
critical energy density.  Details of the lattice discretization, TT projection,
and spectral estimator are collected in Appendix~\ref{app:gravitational-wave-module}.

\subsection{The parameter sets and the snapshots}
\label{sec:results-slices}
In this subsection and the following numerical results, we adopt the following set of {six} parameter choices
\begin{equation}
	\bigl(\alpha,(\beta/H)_n\bigr)
	\in\{(1,5),(1,10),(1,20),(0.2,10),(5,10),{(10,10)}\},
	\label{eq:results-parameter-sample}
\end{equation}
and use $(1,10)$ as the benchmark case. For each parameter choice, we perform five independent simulations. All numerical results quoted below for a given parameter set, including the reported variances, are obtained from these five realizations.

\begin{figure*}[t]
	\centering
	\includegraphics[width=0.96\textwidth]{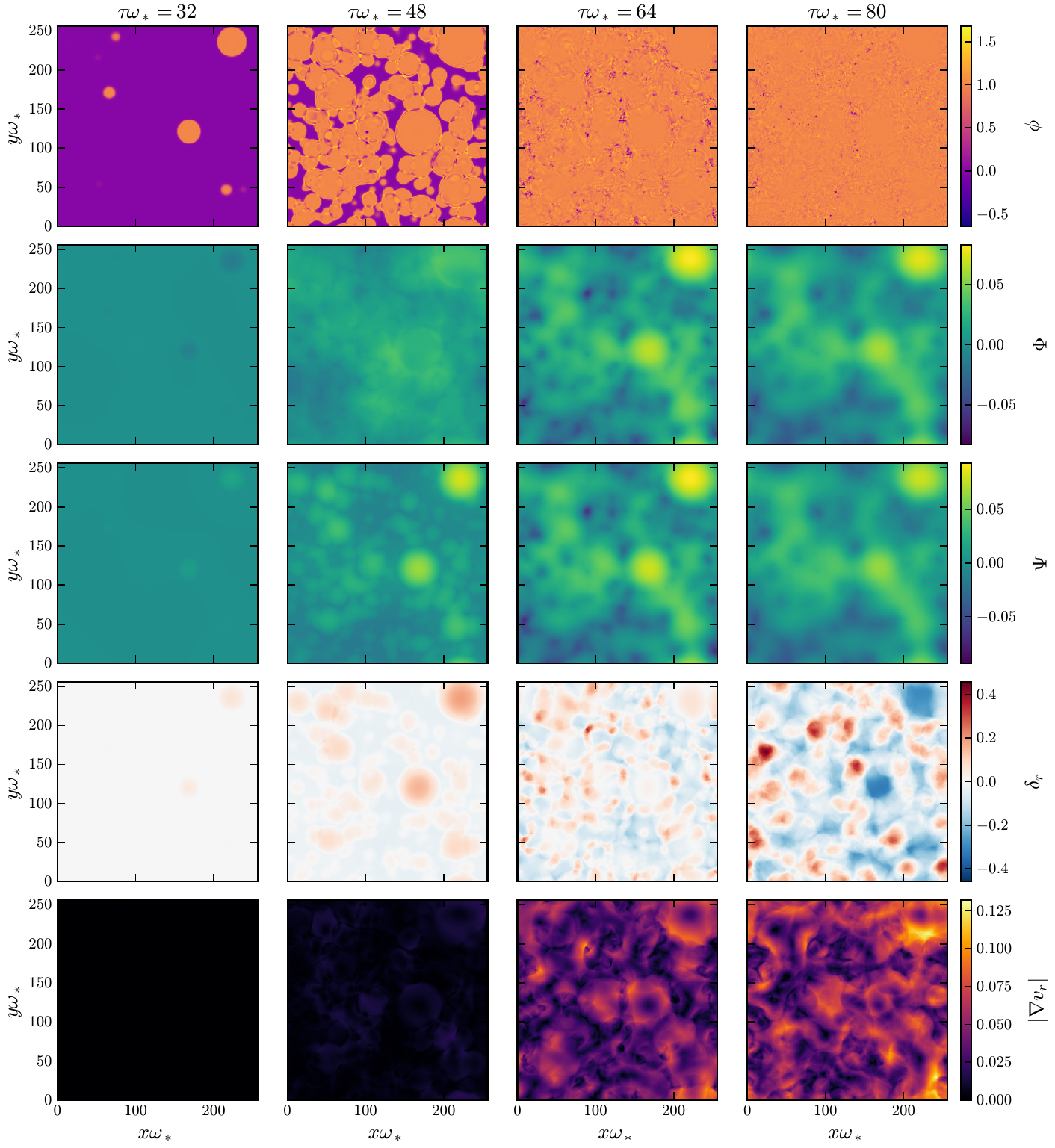}
	\caption{The snapshots of the simulation of FOPT with the parameters $(\alpha,(\beta/H)_n)=(1,10)$.  From top to bottom the panels show the scalar field, the Newtonian potential $\Phi$, spatial-curvature potential $\Psi$, radiation energy density contrast $\delta_r$, and magnitude of the gradient of the longitudinal radiation-velocity potential $|\nabla v_r|$.}
	\label{fig:metric-radiation-slices}
\end{figure*}

Figure~\ref{fig:metric-radiation-slices} shows the snapshots of the
benchmark run, $\alpha=1$, $(\beta/H)_{n}=10$, including the scalar field, the Newtonian
potential $\Phi$, the spatial-curvature potential $\Psi$, the radiation density
contrast $\delta_r$, and the magnitude of the gradient of the longitudinal
radiation-velocity potential $|\nabla v_r|$.
The scalar metric perturbations $\Phi$ and $\Psi$, the radiation energy perturbation, and the longitudinal velocity clearly trace the evolution of the scalar field across the slices. 
{After the phase transition is complete, the scalar metric potentials and radiation density and velocity perturbations persist even as
the scalar field becomes increasingly homogeneous.}
Both $|\Phi|$ and $|\Psi|$ remain roughly an order of magnitude below unity throughout the representative evolution, with $|\Phi|, |\Psi| \sim 0.1 \ll 1$. This justifies our linear approximation for the metric perturbations. Moreover, it also indicates that scalar metric perturbations can indeed be neglected in the scalar field equation of motion compared to the field's self-interactions. Moreover, $\Psi$ and $\Phi$ exhibit a significant difference {during transitions}, which arises from the non-negligible anisotropic stress tensor.

\subsection{The mean value and the variance of the scalar field}
\label{sec:results-sample}

We introduce the parameter $x_\varphi$ to quantify the progress of the FOPT
\begin{equation}
 x_\varphi(\tau)=
 \frac{\langle\varphi\rangle(\tau)-\varphi_{\rm false}}
 {\varphi_{\rm true}-\varphi_{\rm false}},\,
  \rm{with}\,
 \varphi_{\rm false}=0,\,
  \rm{and}\,
 \varphi_{\rm true}=1.
 \label{eq:results-mean-field-clock}
\end{equation}
where $\varphi$ is the dimensionless field value used in the simulation.

\begin{figure}[t]
  \centering
  \includegraphics[width=0.65\linewidth]{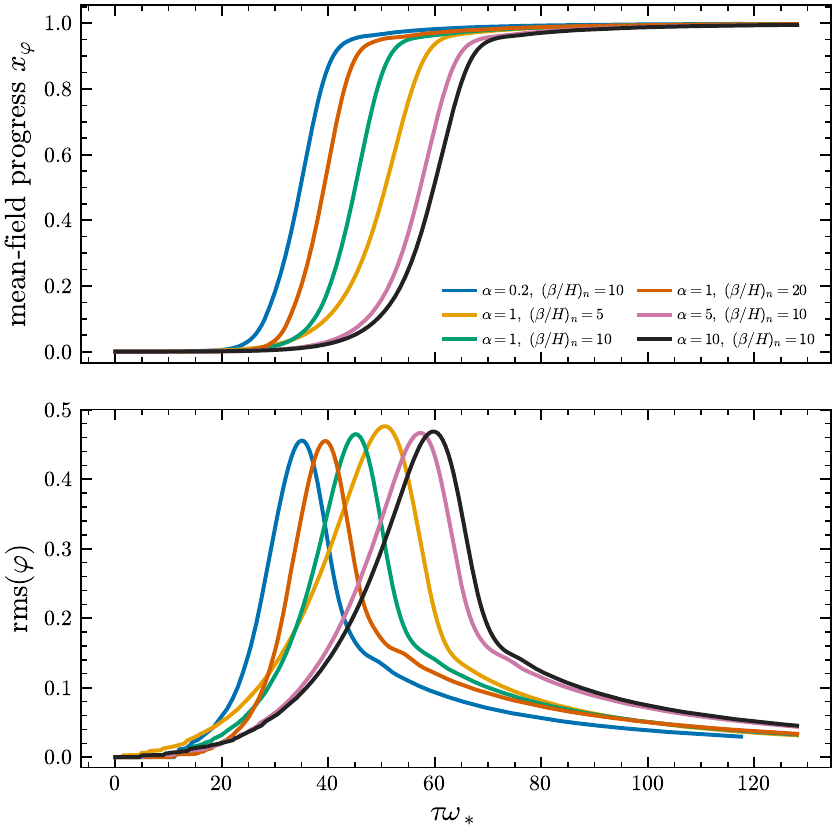}
  \caption{The evolution of $x_{\varphi}$~(top) and $\sigma_{\varphi}$~(bottom) with the parameter sets in Eq~\eqref{eq:results-parameter-sample}.}
  \label{fig:mean-field-evolution}
\end{figure}

Figure~\ref{fig:mean-field-evolution} shows the evolution of $x_\varphi$ and the spatial
standard deviation
\begin{equation}
 \sigma_\varphi=
 \sqrt{\langle\varphi^2\rangle-\langle\varphi\rangle^2}.
 \label{eq:results-scalar-sigma}
\end{equation}
With $\alpha=1$ fixed, we find that increasing $(\beta/H)_n$ causes $x_\varphi$ to reach $1$~(and $\sigma_{\varphi}$ to reach the maximum) more quickly, which implies that the FOPT completes in a shorter interval.  With $(\beta/H)_n=10$ fixed
, {decreasing} $\alpha$ has the same consequence because the smaller
initial radiation density lowers $H_n$ and hence both
$\Gamma(\tau_n)=H_n^4$ and $\beta_{\rm nuc}=(\beta/H)_nH_n$. 
In every run $\sigma_\varphi$ peaks close to
$x_\varphi=1/2$, with
$0.455\leq\sigma_{\varphi,\max}\leq0.478$.  
The nearly universal peak height therefore indicates, for each parameter set, the moment during the FOPT when the true and false vacua have equal volume fractions, while the subsequent slow decay records residual walls and post-collision oscillations.

\subsection{The power spectra of scalar metric perturbations}
\label{sec:results-metric-spectra}
In this subsection, we present rigorous lattice results for the scalar metric perturbation spectra. Compared to the previous semi-analytical results at superhorizon scales~\cite{Franciolini:2025ztf,Wang:2026zvz}, our simulations yield more precise amplitudes, and capture the sub-horizon spectra which are inaccessible to analytical approaches.

To characterize the evolution of the scalar metric perturbations, we define the power spectra of $\Phi$ and $\Psi$, denoted as $P_\Phi$ and $P_\Psi$, according to Eq.~\eqref{eq:setup-power-spectrum}. Note that the presence of the anisotropic stress tensor in our work leads to an non-negligible discrepancy of $\Phi$ and $\Psi$ during their evolution. To quantify this difference, we define the standard signed cross-correlation coefficient~\cite{Cooray:2002dia}
\begin{align}
	r_{\Phi\Psi}(k)=\frac{C_{\Phi\Psi}}{\sqrt{P_\Phi P_\Psi}},\quad C_{\Phi\Psi}(k)=\frac{P_\Phi+P_\Psi-P_{\Phi-\Psi}}{2}.
\end{align}
For nearly equal $\Phi$ and $\Psi$, $r_{\Phi\Psi}$ approaches $1$.

\begin{figure*}[t]
  \centering
  \includegraphics[width=0.8\textwidth]{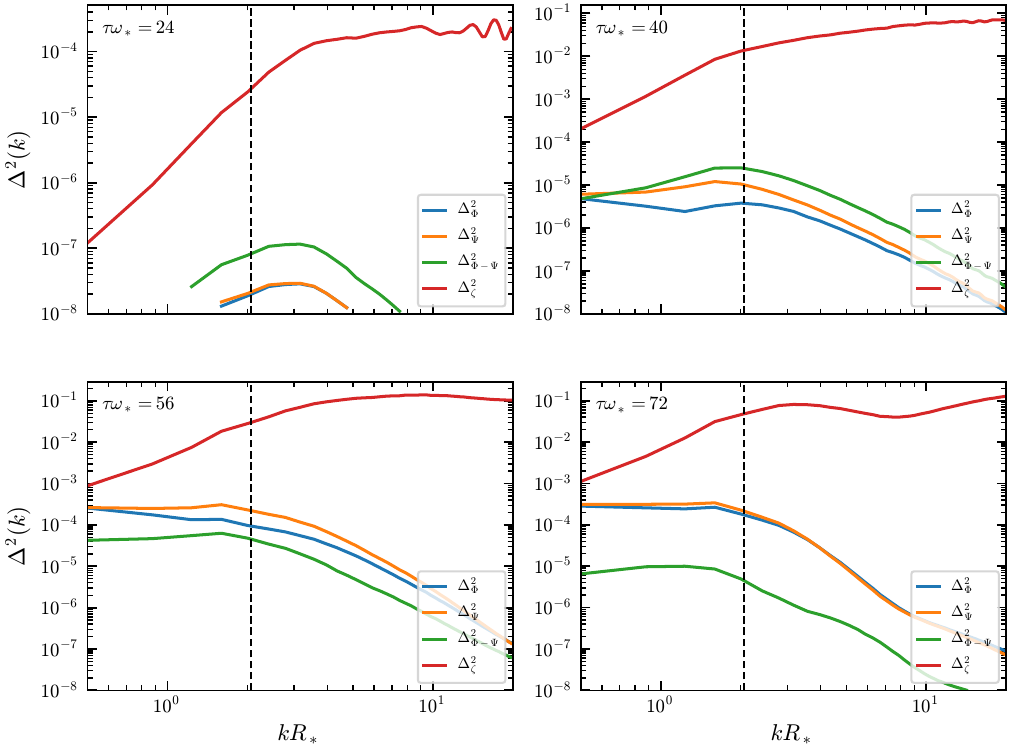}
  \caption{The evolution of the power spectra of $\Phi$, $\Psi$, $\Phi-\Psi$ and the gauge-invariant $\zeta$ at the time points $\omega_{*}\tau=24\,,40\,,56$ and $72$, where the parameters as chosen as $\alpha=1$ and $(\beta/H)_{n}=10$. The vertical dashed lines represent the largest horizon scale in the simulation $k_{\mathcal{H},\rm{max}}=2\pi\mathcal{H}_{\rm{max}}$.
  }
  \label{fig:metric-spectra-times}
\end{figure*}

\begin{figure}[t]
	\centering
	\includegraphics[width=0.65\linewidth]{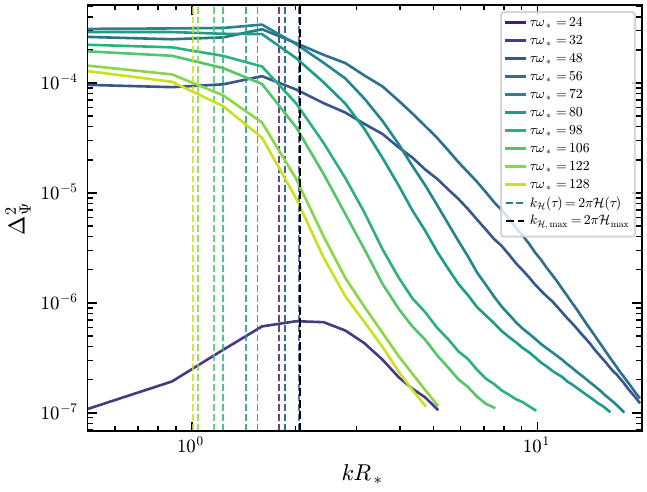}
	\caption{The detailed evolution of the $\Delta_{\Psi}^{2}$ for $\alpha=1$ and $(\beta/H)_{n}=10$.}
	\label{fig:psi-overlay}
\end{figure}

Figure~\ref{fig:metric-spectra-times} shows the evolution of the scalar spectra. Specifically, both $\Phi$ and $\Psi$ are suppressed by $k^{-2}$ on sub-horizon scales and decrease over time within the horizon, as dictated by Eqs.~\eqref{eq:setup-einstein-constraints} and~\eqref{eq:setup-source-combinations}. See also Fig.~\ref{fig:psi-overlay} for more detailed evolution of $\Delta_{\Psi}^{2}$. In contrast, the gauge-invariant variable $\zeta$ is nearly scale-invariant on sub-horizon scales, primarily because it is dominated by energy density perturbations within the horizon, see Eq.~\eqref{eq:setup-zeta}.

\begin{figure}[ht]
  \centering
  \includegraphics[width=0.65\linewidth]{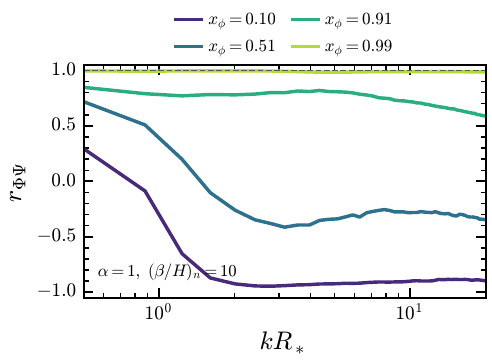}
  \caption{The scale dependence of $r_{\Phi\Psi}$ as a function of the FOPT progress parameter $x_\varphi$ for $\alpha=1$ and $(\beta/H)_{n}=10$.  }
  \label{fig:metric-coherence}
\end{figure}

Figure~\ref{fig:metric-coherence} shows the scale dependence of $r_{\Phi\Psi}$ as a function of the FOPT progress parameter $x_\varphi$. Initially, although the power spectra $P_\Phi$ and $P_\Psi$ are relatively small, the discrepancy between $\Phi$ and $\Psi$ is most pronounced. Across a broad range of scales with $kR_*>1$, $\Phi$ and $\Psi$ have almost opposite signs, leading to $r_{\Phi\Psi}$ approaching $-1$. This behavior originates from the large anisotropic stress generated by the gradient of $\varphi$ in the initial bubbles. As the FOPT proceeds, this difference gradually diminishes and becomes negligible. For instance, at $x_\varphi\simeq0.99$, we obtain $0.977\leq r_{\Phi\Psi}\leq0.996$ across all of the scales.

To characterize the scalar metric perturbation spectra on horizon scales, we average the power spectra over the range $0.5\leq kR_*<2\pi\mathcal H R_*$ at $x_\varphi\simeq0.99$ (when the phase transition is essentially complete), as shown in Table~\ref{tab:metric-ir-amplitudes}. We find that slower phase transitions produce stronger horizon-scale perturbations, which is qualitatively consistent with earlier analytical understandings~\cite{Zou:2025sow,Wang:2026zvz}. Table~\ref{tab:metric-ir-amplitudes} also shows that the effect of $(\beta/H)_n$ is much more pronounced than that of $\alpha$. 
\begin{table}[htpb]
	\centering
	\caption{Power spectra averaged over the range
		$0.5 \leq kR_* < 2\pi \mathcal{H} R_*$ at $x_\varphi \simeq 0.99$.}
	\label{tab:metric-ir-amplitudes}
	\begin{tabular}{ccc}
		\toprule
		$(\alpha,(\beta/H)_n)$
		& $10^4 A_\Phi$
		& $10^4 A_\Psi$ \\
		\midrule
		$(0.2,10)$
		& $0.678 \pm 0.142$
		& $0.770 \pm 0.183$ \\
		
		$(1,5)$
		& $12.4 \pm 4.12$
		& $13.9 \pm 4.12$ \\
		
		$(1,10)$
		& $2.35 \pm 0.179$
		& $2.63 \pm 0.117$ \\
		
		$(1,20)$
		& $0.566 \pm 0.0571$
		& $0.634 \pm 0.0786$ \\
		
		$(5,10)$
		& $4.05 \pm 1.43$
		& $4.88 \pm 1.90$ \\
		
		$(10,10)$
		& $5.64 \pm 0.658$
		& $6.50 \pm 0.652$ \\
		\bottomrule
	\end{tabular}
\end{table}


\subsection{The probability distribution functions 
of energy density perturbations}
\label{sec:results-density-pdf}
In this subsection, we present numerical results for the {PDF} 
of the total energy density perturbation. To obtain the statistical results, we introduce $\delta_{c}^{(R)}$ as the smoothed the total energy density perturbation over regions of radius $R$, see Appendix~\ref{app:pspdf-smoothed-pdf} for details. We also define the standard deviation  
$\delta_{c}^{(R)}$ as
\begin{equation}
 \sigma_\delta^{(R)}={\sqrt{\langle(\delta_{c}^{(R)})^2\rangle},}
 \label{eq:results-density-moments}
\end{equation}
where $\sigma_\delta^{(R)}$ measures the typical fluctuation
amplitude on the smoothing scale $R$.
\cite{Bernardeau:2001qr}.  


In each case, the overdensities are positive for more than half of the regions. 
Consequently, the energy density PDF peaks in the $\delta_{c}^{(R)}>0$ region, which corresponds to a negative skewness. Meanwhile, a negative skewness also implies an enhanced tail in the $\delta_{c}^{(R)}<0$ regime, consistent with the semi-numerical results obtained using stochastic slow-transition treatments in Refs.~\cite{Lewicki:2024ghw,Franciolini:2025ztf}.

\begin{figure*}[t]
  \centering
  \includegraphics[width=0.55\textwidth]{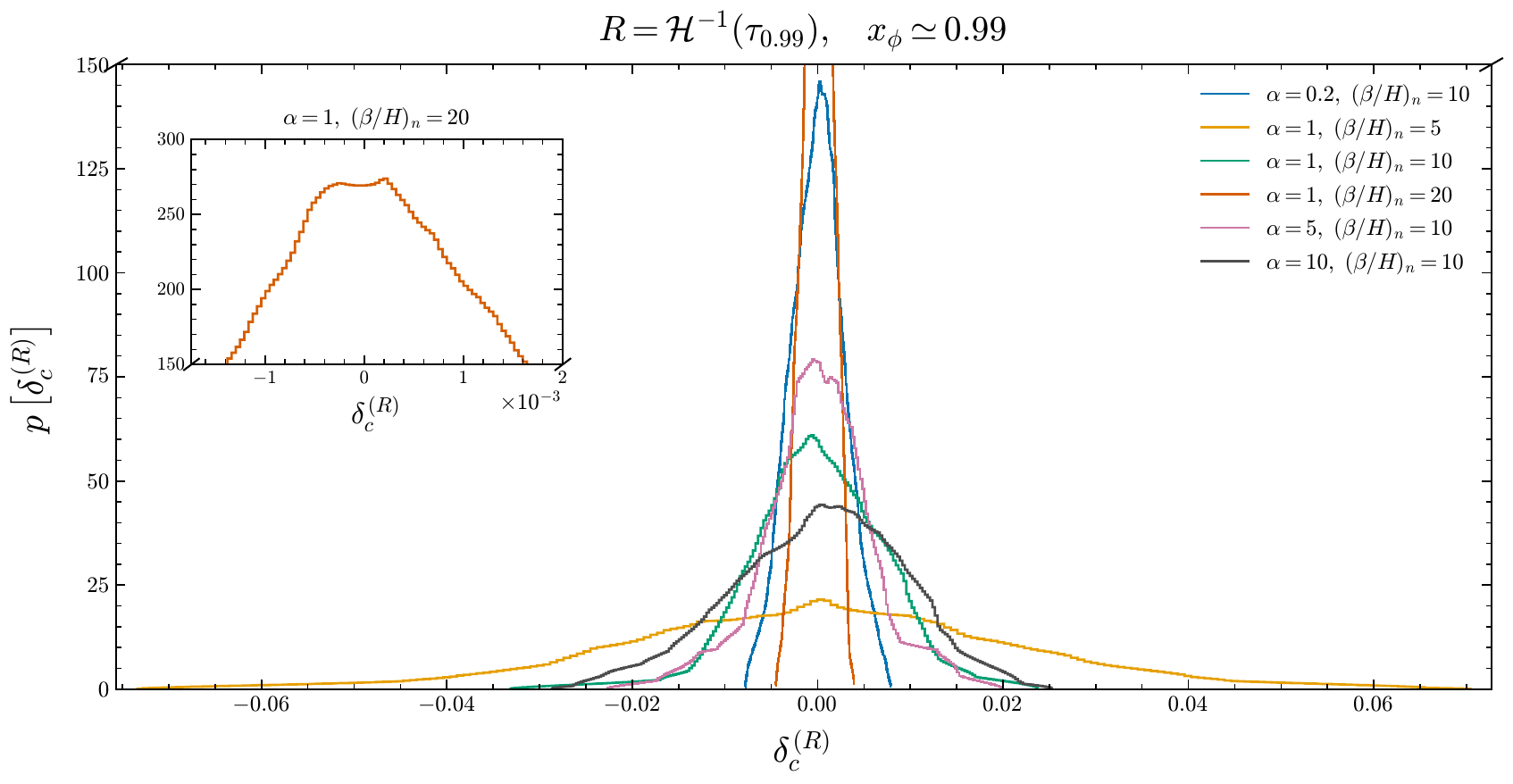}
  \raisebox{0.1cm}{\includegraphics[width=0.38\textwidth]{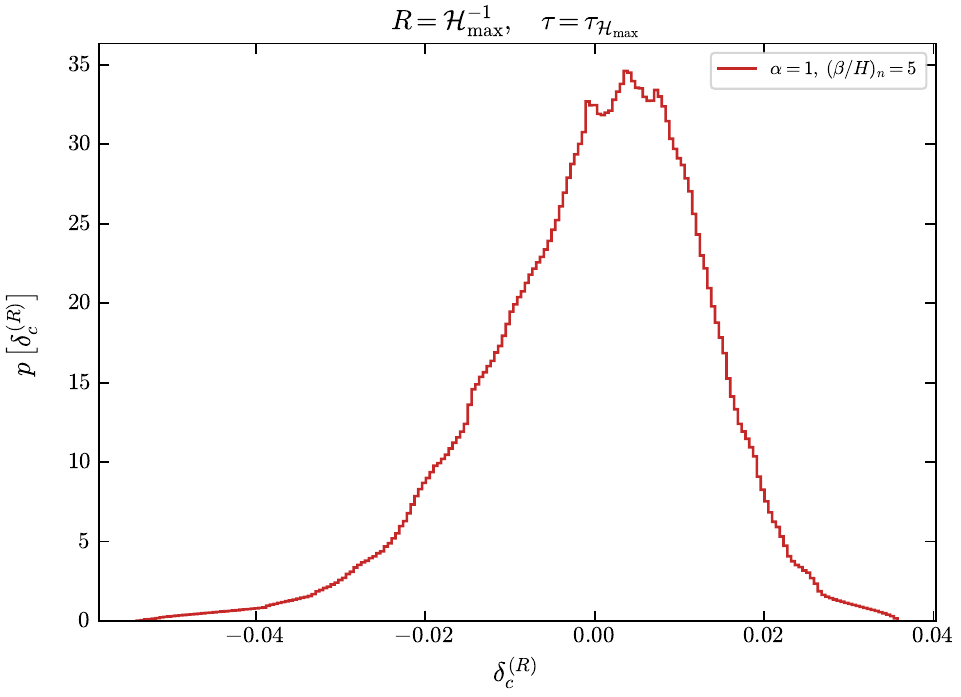}}
  \caption{Probability distributions of the smoothed comoving-density perturbation
  	at the Hubble-horizon scale. Left: comparison among the six parameter
  	sets at $x_\varphi\simeq0.99$, showing that slower transitions generate
  	larger density fluctuations, while the dependence on $\alpha$ is more
  	moderate. Right: the distribution for $\alpha=1$ and
  	$(\beta/H)_n=5$ evaluated when the comoving Hubble horizon reaches its
  	minimum.}
  \label{fig:pdf-parameter-comparison}
\end{figure*}

Table~\ref{tab:density-moments} and the left panel of Fig.~\ref{fig:pdf-parameter-comparison} compare the six parameter sets at
$x_\varphi\simeq0.99$.  {The uncertainties in the table denote the standard error of the mean (SEM).} At fixed $\alpha=1$, the slower FOPT produces
larger horizon-scale perturbations, where $\sigma_{\delta}^{(R)}$ decreases from $2.1\times 10^{-2}$
to $1.5\times 10^{-3}$ as $(\beta/H)_n$ increases from $5$ to $20$. 
At fixed $(\beta/H)_n=10$, increasing $\alpha$ strengthens the
density response more moderately, especially for $\alpha>1$. The right panel of Fig.~\ref{fig:pdf-parameter-comparison} shows the PDF of $\sigma_{\delta}^{(R)}$ at the Hubble-horizon scale for $\alpha=1$ and $(\beta/H)_{n}=5$ when the horizon radius reaches the minimum during the simulation~(at the beginning of the simulation, the comoving Hubble horizon shrinks as the Universe is dominated by vacuum energy, and subsequently grows again once radiation becomes dominant). This figure can be directly compared with the corresponding result in Ref.~{\cite{Franciolini:2025ztf,Wang:2026zvz}}, obtained in the limit $\alpha\to\infty$. Both the standard deviation and the skewness show similar behavior, providing a useful cross-check between the two approaches. This agreement supports the validity of our simulation and the effectiveness of the superhorizon approximation adopted in Ref.~{\cite{Franciolini:2025ztf,Wang:2026zvz}}.

\begin{table}[t]
\centering
\caption{Standard deviation 
of the smoothed comoving-density
PDF at $x_\varphi\simeq0.99$. }
\label{tab:density-moments}
\begin{tabular}{ccc}
\toprule
$\alpha$ & $(\beta/H)_n$ & $\sigma_{{R}}\pm {\mathrm{SEM}}$ \\
\midrule
1 & 5  & \(0.021085\pm0.003529\) \\
1 & 10 &  \(0.006965\pm0.001234\) \\
1 & 20 &  \(0.001485\pm0.000109\)\\
0.2 & 10 & \(0.002786\pm0.000285\) \\
5 & 10 & \(0.005521\pm0.001220\) \\
10 & 10 & \(0.009049\pm0.000916\) \\
\bottomrule
\end{tabular}
\end{table}


\subsection{Gravitational waves}
\label{sec:gw-results}
\begin{figure*}[htbp]
	\centering
	\begin{subfigure}
		\centering
		\includegraphics[width=0.45\textwidth]{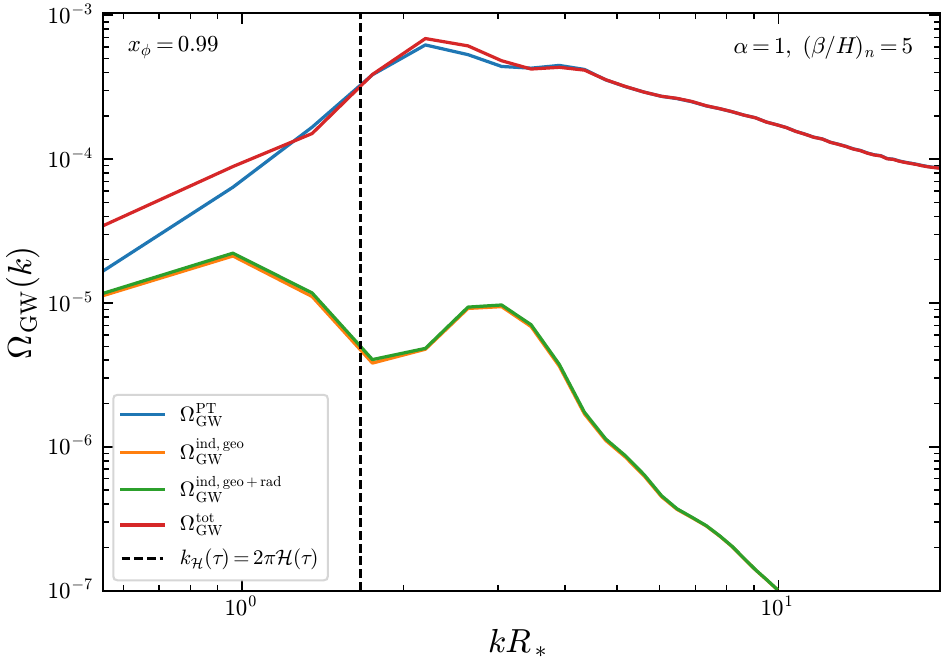}
	\end{subfigure}
	\hfill
	\begin{subfigure}
		\centering
		\includegraphics[width=0.45\textwidth]{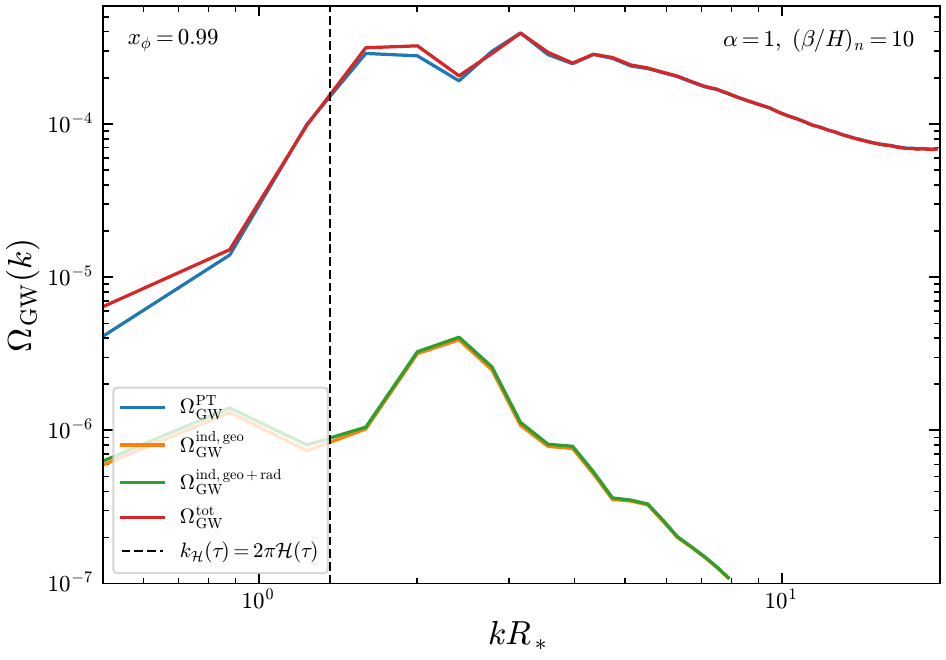}
	\end{subfigure}
		\vspace{0.6cm} 
	\begin{subfigure}
		\centering
		\includegraphics[width=0.45\textwidth]{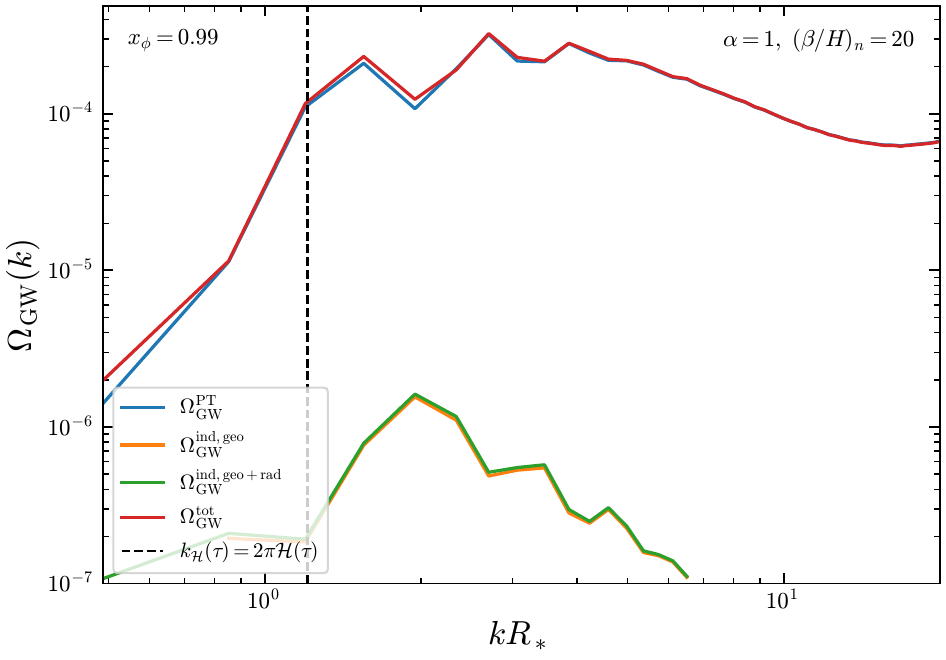}
	\end{subfigure}
    \hfill
	\begin{subfigure}
		\centering
		\includegraphics[width=0.45\textwidth]{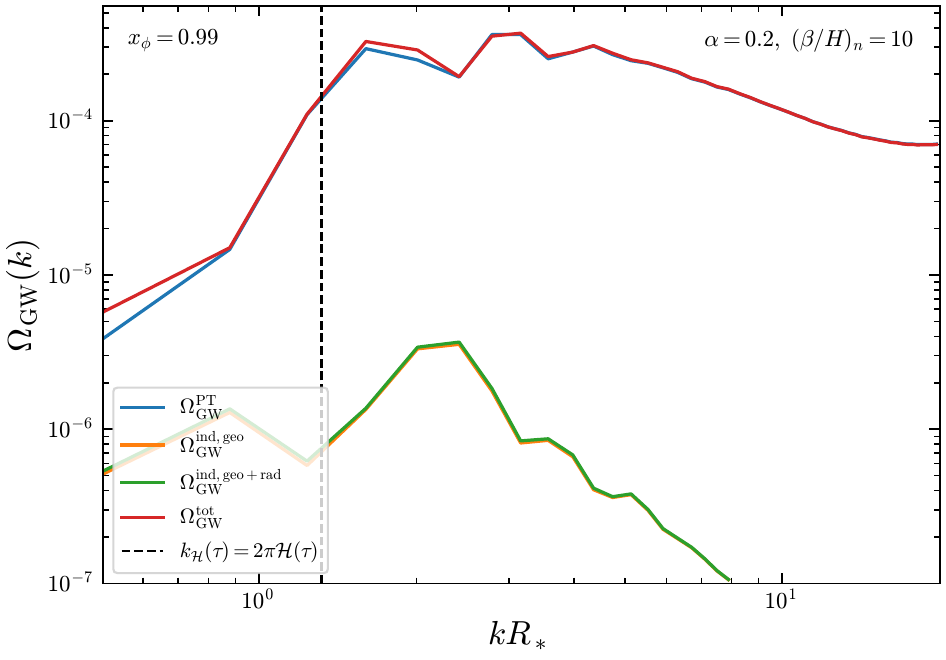}
	\end{subfigure}
    	\vspace{0.6cm} 
	\begin{subfigure}
		\centering
		\includegraphics[width=0.45\textwidth]{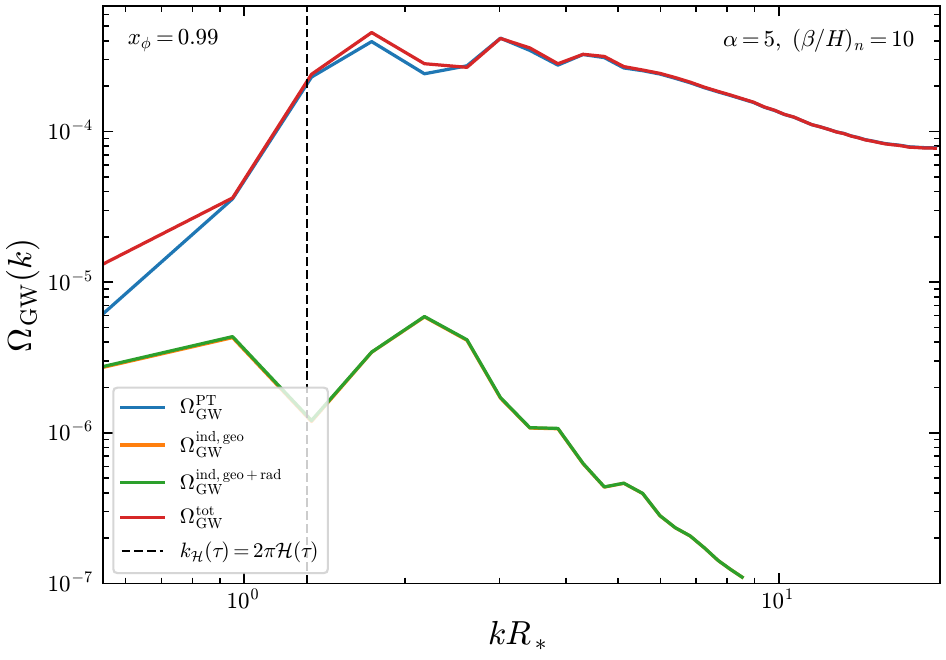}
	\end{subfigure}	
    \hfill
    	\begin{subfigure}
		\centering
		\includegraphics[width=0.45\textwidth]{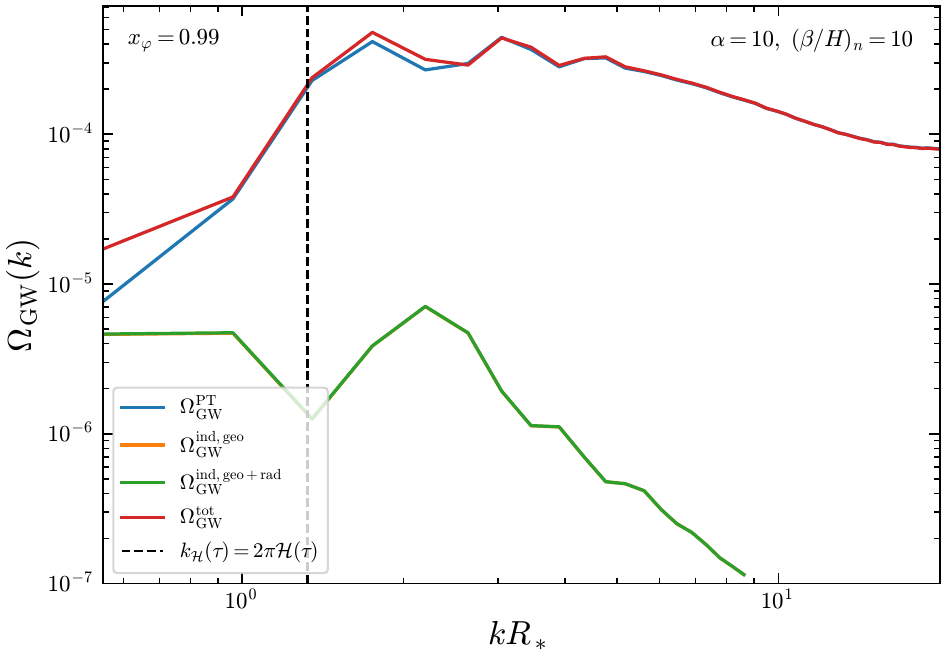}
	\end{subfigure}	
	\caption{GW energy spectra from the different source channels for the six parameter
		sets, evaluated at $x_{\varphi}=0.99$. The blue curves show the spectra
		sourced by the scalar field, the orange curves correspond to the
		metric-induced contribution, the green curves include both the metric and
		radiation-velocity sources, and the red curves give the total spectra obtained
		by combining all three contributions.}
	\label{fig:gw-channel-comparison}
\end{figure*}

We now turn to GW production and compare the standard phase-transition contribution sourced by the scalar field with the additional tensor modes generated by the scalar metric and radiation perturbations. Figure~\ref{fig:gw-channel-comparison} shows the GW energy spectra from each source which are evaluated at $x_{\varphi}=0.99$ for each parameter set. The blue curves show the GW spectrum sourced by the scalar field, while the orange curves correspond to the metric-induced contribution. The green curve includes both the metric and radiation-velocity sources, and the red curve gives the total spectrum obtained by combining all three contributions.
The vertical dashed line indicates the Hubble scale at the corresponding time. On subhorizon scales, the GW energy spectrum sourced by the scalar field nearly coincides with the total spectrum, indicating that the additional induced contributions remain subdominant in this regime. The close agreement between the orange and green curves reflects the fact that the radiation perturbations are themselves induced by the scalar metric perturbations and therefore provide only a secondary correction to the GW spectrum.
At the superhorizon scales, the contributions from sources other than the scalar field become increasingly important, modifying the infrared power-law behavior and leading to a scale-dependent spectral index.

To quantify the enhancement of the GW signal due to the additional
metric and radiation sources, we integrate the GW energy spectrum over
a given wavenumber range,
\begin{align}
	\Omega_X^{\rm band}(\tau)
	&=
	\int_{k_{\rm min}}^{k_{\rm max}}
	\Omega_X(k,\tau)\,d\ln k ,
	\nonumber\\
	\epsilon(\tau)
	&=
	\frac{
		\Omega_{\rm tot}^{\rm band}(\tau)
		-
		\Omega_{\rm PT}^{\rm band}(\tau)}
	{\Omega_{\rm PT}^{\rm band}(\tau)} ,
	\label{eq:results-gw-enhancement}
\end{align}
where $\epsilon$ measures the enhancement of the total GW energy relative
to that sourced directly by the scalar field. We consider both the full
resolved range, $0.5/R_*<k<20/R_*$, and the long-wavelength range
$k<2\pi\mathcal H$.

At the common conversion stage $x_\varphi\simeq0.99$, the resulting
enhancements are summarized in Table~\ref{tab:gw-enhancement}. Over the
full resolved band, $\epsilon$ is a few percent, reaching
$(4.73\pm0.42)\%$ for the slowest transition with $(\beta/H)_n=5$.
The enhancement becomes more pronounced toward the Hubble and
super-Hubble scales, where it reaches $(9.77\pm3.11)\%$ for the same
case. This confirms that the metric and radiation contributions are
relatively more important at long wavelengths, despite giving only a
modest correction to the integrated GW energy over the full spectrum.
Their impact is therefore most significant in the infrared, where they
also modify the spectral behavior of the total GW signal.
\begin{table}[htbp]
	\centering
	\caption{Enhancement of the total GW energy relative to the direct
		scalar-field contribution, integrated over the full resolved band and
		over $k<2\pi\mathcal H$.}
	\label{tab:gw-enhancement}
	\begin{tabular}{cccc}
		\toprule
		$\alpha$ & $(\beta/H)_n$
		& $\epsilon_{0.5/R_*<k<20/R_*}$ (\%)
		& $\epsilon_{k<2\pi\mathcal H}$ (\%) \\
		\midrule
		1 & 5  & $4.73\pm0.42$   & $9.77\pm3.11$ \\
		1 & 10 & $3.430\pm0.082$ & $3.42\pm0.56$ \\
		1 & 20 & $2.989\pm0.093$ & $4.69\pm0.45$ \\
		\bottomrule
	\end{tabular}
\end{table}

Patch-based analyses of slow transitions follow super-Hubble stochastic histories and estimate the induced signal from curvature correlators or convolution kernels~\cite{Franciolini:2025ztf,Wang:2026zvz}. In contrast, our three-dimensional simulation directly evolves the local sources over the full range of scales around the Hubble horizon, without assuming that the superhorizon perturbations remain frozen until horizon entry. We find that the near-horizon contribution is larger than suggested by such semi-analytic estimates, leading to a several-percent enhancement of the total GW spectrum on super-Hubble scales. The effect is particularly important in the infrared, where it modifies the spectral behavior and the effective power-law index.

\section{Conclusion and discussion}
\label{sec:discussion-conclusions}
We conduct the first three-dimensional multi-bubble lattice simulation of a FOPT that self-consistently evolves the scalar field, scalar metric perturbations, and the energy density and velocity perturbations of the background radiation. This framework also allows us to comprehensively characterize the resulting GW spectrum sourced by all these components. The simulation is performed in the Newtonian gauge, where the scalar metric perturbations are described by the Newtonian potential $\Phi$ and the spatial-curvature potential $\Psi$, determined by the scalar sector of the Einstein equations. We assume that the scalar field and the background radiation interact only gravitationally. Our simulations simultaneously resolve the bubble walls and retain information up to the horizon scale. Compared with previous semi-numerical treatments focused mainly on superhorizon scales, our fully numerical calculation provides a continuous and more precise description of the perturbation spectrum, particularly on subhorizon and near-horizon scales. We further obtain the PDF of the comoving-gauge energy-density perturbation on the horizon scale. The PDF exhibits negative skewness, with its peak shifted to $\delta_c>0$ and an enhanced negative tail, in agreement with previous semi-analytical results.

Our simulations indicate that horizon-scale perturbation amplitudes are far below the PBH formation threshold, even in the case of $\alpha=1, (\beta/H)_n=5$. However, in highly rare scenarios where tunneling is substantially delayed within the horizon, PBHs could potentially form due to a combination of a lower internal equation-of-state parameter and the accumulation of incoming bubble wall energy. Resolving such anomalous regions lies beyond the reach of our current simulation capabilities.

{Beyond the possible collapse of rare overdense regions, the tendency toward negative skewness motivates further investigation of the underdense regions produced by the transition. Although one-point density statistics do not determine their spatial extent or connectivity, reconstructing their profiles and following their nonlinear evolution would clarify whether the transition can seed void-like configurations~\cite{Joana:2025gqf}. A more complete account of the subsequent evolution would also require full fluid dynamics and scalar-field–fluid interactions. Extending the present framework in this direction would allow us to study both the redistribution of density perturbations by the fluid and the gravitational waves generated by sound waves, including their interplay with the direct scalar-field and scalar-induced contributions.}
\ifdefined\INTEGRATEDMANUSCRIPT
\else
\documentclass[11pt,a4paper]{article}

\usepackage[T1]{fontenc}
\usepackage{lmodern}
\usepackage[a4paper,margin=25mm]{geometry}
\usepackage{amsmath}
\usepackage{amssymb}
\usepackage{bm}
\usepackage{cite}
\usepackage{xcolor}
\usepackage{hyperref}
\newcommand{\rev}[1]{#1}

\hypersetup{
  colorlinks=true,
  linkcolor=blue,
  citecolor=blue,
  urlcolor=blue,
  pdftitle={Appendix on Numerical Methods and Statistical Estimators},
  pdfauthor={}
}

\setlength{\emergencystretch}{2em}
\setlength{\parindent}{2em}
\allowdisplaybreaks

\begin{document}
\fi

\section*{Appendices}
\appendix
\ifdefined\PRDMANUSCRIPT
\else
\numberwithin{equation}{section}
\fi
The appendices detail the numerical treatment of exponential bubble nucleation,
constraint-projected scalar metric and radiation perturbations, direct and
scalar-induced tensor modes, and the associated spectral and one-point estimators.
These elements form a unified computational framework: nucleation generates scalar-field
inhomogeneities; the Einstein constraints determine \(\Phi\), \(\Psi\), and the
radiation response; and the resulting matter and scalar sources drive the tensor
channels.  Power spectra and real-space probability distributions are then
constructed from the same evolved fields.  Relevant physical background can be found in
Refs.~\cite{Turner:1992tz,Ma:1995ey,Baumann:2007zm,Figueroa:2020rrl}.

\section{Exponential bubble-nucleation prescription}
\label{app:exponential-nucleation-module}

This section presents the exponential bubble-nucleation prescription used in our
simulations of a {FOPT}.  The simulation evolves in
conformal time \(\tau\), whereas the nucleation rate itself is defined in physical
spacetime.  The prescription therefore first specifies the physical nucleation
rate \(\Gamma_{\rm phys}\), converts it into an inhomogeneous Poisson process in
conformal time, and injects a bubble profile into the scalar field \(\phi\) at
each accepted event \cite{Turner:1992tz}.

\subsection{Background Evolution and Nucleation Rate}
\label{app:nucleation-background-rate}

This section describes the background evolution and nucleation rate used by the nucleation prescription. The program uses the dimensionless conformal time \(\widehat\tau\) as the evolution variable, while the exponential growth of the physical nucleation rate is described in terms of the dimensionless physical time \(\widehat t\). They are related by
\begin{equation}
  \mathrm{d}\widehat t
  =
  a(\widehat\tau)\,\mathrm{d}\widehat\tau ,
  \label{eq:nuc-dimensionless-physical-time}
\end{equation}
where \(a(\widehat\tau)\) is the scale factor. We denote the reference nucleation time by \(\widehat\tau_n\) and the initial time by \(\widehat\tau_i\). The corresponding dimensionless reference physical time for nucleation is
\begin{equation}
  \widehat t_n
  \equiv
  \widehat t(\widehat\tau_n)
  =
  \int_{\widehat\tau_i}^{\widehat\tau_n}
  a(\widehat\tau)\,\mathrm{d}\widehat\tau .
  \label{eq:nuc-tn}
\end{equation}

The phase-transition strength \(\alpha\) is defined as the ratio between the vacuum energy density and the radiation energy density at the reference nucleation time,
\begin{equation}
  \alpha
  =
  \frac{V_0}{\bar\rho_r(\widehat\tau_n)} ,
  \label{eq:nuc-alpha-definition}
\end{equation}
where \(V_0\) denotes the vacuum energy density, and \(\bar\rho_r\) denotes the homogeneous radiation background energy density. The radiation background redshifts as
\begin{equation}
  \bar\rho_r(\widehat\tau)
  =
  \frac{\rho_{r0}}{a^4(\widehat\tau)}
  \label{eq:nuc-radiation-redshift}
\end{equation}
where we set \(a(\widehat\tau_i)=1\), so that \(\rho_{r0}\) is the initial radiation background energy density. Equations~\eqref{eq:nuc-alpha-definition} and \eqref{eq:nuc-radiation-redshift} imply that \(\rho_{r0}\) must satisfy
\begin{equation}
  \frac{\rho_{r0}}{a_n^4}
  =
  \frac{V_0}{\alpha},
  \qquad
  a_n\equiv a(\widehat\tau_n).
  \label{eq:nuc-rho-r0-condition}
\end{equation}
Since \(a_n\) itself depends on the background evolution, \(\rho_{r0}\) cannot be obtained directly from Eq.~\eqref{eq:nuc-rho-r0-condition} as an algebraic input; it must instead be solved self-consistently together with the Friedmann equation.

In the conformal-time program variables used here, the background scale factor obeys
\begin{equation}
  \left(
    \frac{\mathrm{d}a}{\mathrm{d}\widehat\tau}
  \right)^2
  =
  \frac{1}{3}
  \left(\frac{f_*}{M_{\rm Pl}}\right)^2
  \left(
    V_0 a^4+\rho_{r0}
  \right),
  \label{eq:nuc-background-friedmann}
\end{equation}
where \(f_*\) is the model energy scale and \(M_{\rm Pl}\) is the reduced Planck mass. The program adjusts \(\rho_{r0}\) and integrates from \(\widehat\tau_i\) to \(\widehat\tau_n\) until Eq.~\eqref{eq:nuc-rho-r0-condition} is satisfied. The resulting \(a(\widehat\tau)\) determines both the reference physical time \(\widehat t_n\) and the background evolution prior to the reference time.

The physical Hubble parameter at the reference time is fixed by the total background energy density. Using \(\bar\rho_r(\widehat\tau_n)=V_0/\alpha\), one obtains
\begin{equation}
  \widehat H_n
  =
  \frac{f_*}{M_{\rm Pl}}
  \left[
    \frac{V_0}{3}
    \left(1+\alpha^{-1}\right)
  \right]^{1/2},
  \label{eq:nuc-hubble-anchor}
\end{equation}
where \(\widehat H_n\) is the dimensionless physical Hubble parameter at the reference transition time. Thus, once \(V_0\) and \(\alpha\) are specified, \(\widehat H_n\) is fixed by the energy density at the reference time. The quantities that still have to be determined from the background integration for the nucleation rate are the reference physical time \(\widehat t_n\) and the scale factor \(a(\widehat\tau)\) entering the comoving nucleation rate.

The physical nucleation rate is normalized at the reference time as
\begin{equation}
  \widehat\Gamma_n
  =
  \widehat H_n^4,
  \qquad
  \widehat\beta_{\rm phys}
  =
  \left(\frac{\beta}{H}\right)_n
  \widehat H_n,
  \label{eq:nuc-rate-normalization}
\end{equation}
where \((\beta/H)_n\) is the input parameter controlling the growth rate of the nucleation rate, and the subscript \(n\) denotes evaluation at the reference nucleation time. The nucleation rate per unit physical volume and per unit physical time is then
\begin{equation}
  \widehat\Gamma_{\rm phys}(\widehat t)
  =
  \widehat\Gamma_n
  \exp\!\left[
    \widehat\beta_{\rm phys}
    \left(
      \widehat t-\widehat t_n
    \right)
  \right].
  \label{eq:nuc-physical-rate}
\end{equation}
The nucleation rate per unit comoving volume and per unit conformal time is
\begin{equation}
  \widehat\Gamma_{\tau}(\widehat\tau)
  =
  a^4(\widehat\tau)\,
  \widehat\Gamma_{\rm phys}
  \bigl(\widehat t(\widehat\tau)\bigr).
  \label{eq:nuc-conformal-rate}
\end{equation}

When the evolution first crosses \(\widehat\tau_n\), the module updates the anchor values using the actual evolved values of \(a(\widehat\tau_n)\) and \(\widehat t_n\). Only a small number of bubbles are generated before \(\widehat\tau_n\) and the background evolution is not significantly modified, this online update only produces differences within the numerical integration and interpolation errors.

\subsection{Candidate Event Generation and Bubble Injection}
\label{app:nucleation-events-and-injection}

For a simulation box with comoving volume \(V_{\rm com}\), the instantaneous event intensity \(\lambda\) is
\begin{equation}
  \lambda(\widehat\tau)
  =
  V_{\rm com}\,
  \widehat\Gamma_\tau(\widehat\tau),
  \label{eq:nuc-lambda}
\end{equation}
which represents the nucleation probability per unit conformal time in the whole box. In a time step \([\widehat\tau_a,\widehat\tau_a+\Delta\widehat\tau]\), the expected number of candidate events \(\mu\) is
\begin{equation}
  \mu(\widehat\tau_a,\widehat\tau_a+\Delta\widehat\tau)
  =
  \int_{\widehat\tau_a}^{\widehat\tau_a+\Delta\widehat\tau}
  \lambda(\widehat\tau)\,\mathrm{d}\widehat\tau ,
  \label{eq:nuc-mu}
\end{equation}
where \(\Delta\widehat\tau\) is the time-step size. In the implementation, this integral is estimated in each time step using Simpson's rule.

Candidate events are generated as an inhomogeneous Poisson process. If \(u_j\in(0,1)\) is the \(j\)-th uniform random number, the waiting increment in integrated intensity is
\begin{equation}
  \Delta I_j=-\ln u_j .
  \label{eq:nuc-waiting-intensity}
\end{equation}
The candidate nucleation time \(\widehat\tau_j\) of the \(j\)-th event is determined by
\begin{equation}
  \int_{\widehat\tau_{\rm c}}^{\widehat\tau_j}
  \lambda(\widehat\tau)\,\mathrm{d}\widehat\tau
  =
  \Delta I_j
  \label{eq:nuc-invert-intensity}
\end{equation}
where \(\widehat\tau_{\rm c}\) is the current cursor position of the event catalog. This equation is inverted by bisection. Candidate centers are sampled uniformly in the periodic simulation box.
{This construction retains the continuous exponential variation of
the rate within a time step.}

At late times, the rapidly increasing nucleation rate can produce many candidate bubble centers within a single time step, while the probability for a candidate to remain in the false-vacuum region becomes very small. Processing all such candidates would therefore add substantial computational cost with little effect on the accepted bubble population. We thus limit the number of candidate centers examined in each time step to 50, which is sufficient to sample the residual false-vacuum regions and provides an efficient stopping mechanism as the transition approaches completion.

After a candidate event is generated, the module first checks whether the neighborhood of the candidate center is already on the true-vacuum side. Let \(\mathcal B_c(R_c)\) denote the sphere centered at the candidate center \(\mathbf{x}_c\) with physical radius \(R_c\), let \(V_{R_c}\) be the volume of this sphere, and let \(\mathbf{x}\) denote the spatial position. The corresponding averaged scalar field \(\bar{\phi}_{R_c}\) is
\begin{equation}
  \bar{\phi}_{R_c}
  =
  \frac{1}{V_{R_c}}
  \int_{\mathcal B_c(R_c)}
  \phi(\mathbf{x})\,\mathrm{d}^3x .
  \label{eq:nuc-local-average}
\end{equation}
If this average has crossed the threshold \(\phi_{\rm th}\),
\begin{equation}
  \phi_{\rm th}
  =
  \frac{1}{2}
  \left(\phi_{\rm false}+\phi_{\rm true}\right),
  \label{eq:nuc-threshold}
\end{equation}
where \(\phi_{\rm false}\) and \(\phi_{\rm true}\) are the false- and true-vacuum field values, respectively, the candidate is rejected in order to avoid repeated nucleation in regions that have already transitioned.

An accepted event injects a radial profile as a function of the physical distance \(r_{\rm phys}\), where \(r_{\rm com}\) is the comoving distance:
\begin{equation}
  r_{\rm phys}=a(\tau_j)r_{\rm com}.
\end{equation}
The profile function \(\phi_{\rm prof}\) is taken to be
\begin{equation}
  \begin{aligned}
  \phi_{\rm prof}(r_{\rm phys})
  &=
  \frac{\Delta\phi}{2}
  \left[
    1-\tanh\!\left(
      \frac{r_{\rm phys}-R_c}{\ell_0}
    \right)
  \right], \\
  \Delta\phi
  &=\phi_{\rm true}-\phi_{\rm false}.
  \end{aligned}
  \label{eq:nuc-profile}
\end{equation}
Here \(\Delta\phi\) is the difference between the true- and false-vacuum field values, \(\ell_0\) is the wall thickness of the critical bubble, and \(R_c\) is the radius of the critical bubble. The critical-bubble profile is obtained from FindBounce~\cite{Guada:2020xnz}. The field variable is modified only within the region \(r_{\rm phys}\le 2R_c\). We use the field update
\begin{equation}
  \phi_{\rm new}
  =
  \sqrt{
    \phi_{\rm old}^2
    +
    \phi_{\rm prof}^2
  },
  \label{eq:nuc-field-update}
\end{equation}
where \(\phi_{\rm old}\) and \(\phi_{\rm new}\) denote the field values before and after injection, respectively.
{This quadrature prescription regulates overlapping injections and
defines the field update used in the simulations.}

\section{Evolution of Scalar Metric and Radiation Perturbations}
\label{app:curvature-perturbation-module}

This appendix specifies the evolution prescription for scalar metric and
radiation perturbations used in our numerical simulations. We work in the
conformal Newtonian gauge,
\begin{equation}
  ds^2
  =
  a^2(\tau)
  \left[
    -(1+2\Phi)d\tau^2
    +(1-2\Psi)\delta_{ij}dx^i dx^j
  \right],
  \label{eq:curv-newtonian-gauge}
\end{equation}
where \(\tau\) is conformal time, a prime denotes differentiation with
respect to \(\tau\), and
\begin{equation}
  \mathcal H\equiv\frac{a'}{a}.
  \label{eq:curv-conformal-H}
\end{equation}
Here \(\Phi\) is the lapse potential and \(\Psi\) is the spatial-curvature
potential. Since gradients of an inhomogeneous scalar field can generate
anisotropic stress, the gravitational slip \(\Phi-\Psi\) is retained in
general.

The numerical formulation is formulated for a three-dimensional model with a
single real scalar field. The scalar field itself is evolved using the
{background Friedmann--Robertson--Walker lattice equations}. At each time, the metric-perturbation module
treats the inhomogeneous scalar-field configuration as an external source and
solves the linearized Einstein equations for \(\Phi\), \(\Psi\), and
\(\Psi'\).

\subsection{Scalar-field sources and radiation perturbations}
\label{app:curv-sources}

Let \(\bar\rho_r\) denote the background radiation energy density. The linear
radiation perturbations are described by
\begin{align}
  \delta_r
  \equiv
  \frac{\delta\rho_r}{\bar\rho_r},\quad
  \theta_r
  \equiv
  \partial_i v_r^i .
  \label{eq:curv-radiation-vars}
\end{align}
The scalar-field sources are constructed from the instantaneous lattice field
\(\phi\) and its conjugate momentum \(\pi_\phi\). With the normalization used
in the code, the energy-density and pressure perturbations are
\begin{align}
	\delta\rho_\phi
	&=
	\frac{1}{2a^6}\pi_\phi^2
	+
	\frac{1}{2a^2}|\nabla\phi|^2
	\nonumber
	+V(\phi)
	-
	\left\langle
	\frac{1}{2a^6}\pi_\phi^2
	+
	\frac{1}{2a^2}|\nabla\phi|^2
	+
	V(\phi)
	\right\rangle ,
	\\
	\delta p_\phi
	&=
	\frac{1}{2a^6}\pi_\phi^2
	-
	\frac{1}{6a^2}|\nabla\phi|^2
	-V(\phi)
	-
	\left\langle
	\frac{1}{2a^6}\pi_\phi^2
	-
	\frac{1}{6a^2}|\nabla\phi|^2
	-
	V(\phi)
	\right\rangle .
	\label{eq:curv-delta-p-phi}
\end{align}
Subtracting the spatial average is equivalent to removing the corresponding
Fourier zero mode. Accordingly, all subsequent \(1/k^2\) projections are
applied only to nonzero Fourier modes.

The scalar-field momentum source is described by the scalar potential
\(q_{\phi,\mathbf k}\), which is obtained by longitudinally projecting the
mixed momentum-density component
\begin{equation}
  T^0{}_{i,\phi}
  =
  -a^{-4}\pi_\phi\,\partial_i\phi.
  \label{eq:curv-momentum-density}
\end{equation}
Explicitly,
\begin{equation}
  q_{\phi,\mathbf k}
  =
  \frac{a^2}{k^2}\,
  i k_i
  \left(T^0{}_{i,\phi}\right)_{\mathbf k}.
  \label{eq:curv-qphi}
\end{equation}
The anisotropic stress is supplied by the traceless part of the scalar-field
gradient tensor. Using mixed spatial indices,
\begin{equation}
  \Sigma^i{}_{j,\phi}
  =
  a^{-2}
  \left(
    \partial^i\phi\,\partial_j\phi
    -
    \frac13\delta^i{}_j|\nabla\phi|^2
  \right),
  \label{eq:curv-sigma-phi}
\end{equation}
and its scalar projection is
\begin{equation}
  \Pi^S_{\phi,\mathbf k}
  =
  \frac32
  \left(
    \frac{k_i k^j}{k^2}
    -
    \frac13\delta_i{}^j
  \right)
  \left(\Sigma^i{}_{j,\phi}\right)_{\mathbf k}.
  \label{eq:curv-pis-projection}
\end{equation}

\subsection{Einstein-constraint projection and radiation-fluid evolution}
\label{app:curv-projection-evolution}

For each nonzero Fourier mode, we define the three source combinations
\begin{align}
  B_{\mathbf k}
  &\equiv
  4\pi G a^2
  \left(
    \delta\rho_{\phi,\mathbf k}
    +
    \bar\rho_r\delta_{r,\mathbf k}
  \right),
  \label{eq:curv-Rk} \\
  M_{\mathbf k}
  &\equiv
  4\pi G
  \left[
    q_{\phi,\mathbf k}
    +
    \frac{4a^2\bar\rho_r}{3k^2}\theta_{r,\mathbf k}
  \right],
  \label{eq:curv-Mk} \\
  A_{\mathbf k}
  &\equiv
  8\pi G a^2\Pi^S_{\phi,\mathbf k}.
  \label{eq:curv-Ak}
\end{align}
The \(00\), \(0i\), and traceless spatial components of the linearized
Einstein equations yield, respectively,
\begin{align}
  -k^2\Psi_{\mathbf k}
  -
  3\mathcal H
  \left(
    \Psi'_{\mathbf k}
    +
    \mathcal H\Phi_{\mathbf k}
  \right)
  &=
  B_{\mathbf k},
  \label{eq:curv-00-constraint} \\
  \Psi'_{\mathbf k}
  +
  \mathcal H\Phi_{\mathbf k}
  &=
  M_{\mathbf k},
  \label{eq:curv-momentum-constraint} \\
  k^2(\Phi_{\mathbf k}-\Psi_{\mathbf k})
  &=
  A_{\mathbf k}.
  \label{eq:curv-anisotropic-constraint}
\end{align}
We impose these constraints through a projection at every time step. Rather
than integrating \(\Psi_{\mathbf k}\) as an independent dynamical variable,
we reconstruct the metric perturbations directly from the instantaneous total
sources. Equations~\eqref{eq:curv-00-constraint}--\eqref{eq:curv-anisotropic-constraint}
therefore give
\begin{align}
  \Psi_{\mathbf k}
  &=
  -
  \frac{
    B_{\mathbf k}
    +
    3\mathcal H M_{\mathbf k}
  }{k^2},
  \label{eq:curv-phi-projection} \\
  \Phi_{\mathbf k}
  &=
  \Psi_{\mathbf k}
  +
  \frac{A_{\mathbf k}}{k^2},
  \label{eq:curv-psi-projection} \\
  \Psi'_{\mathbf k}
  &=
  M_{\mathbf k}
  -
  \mathcal H\Phi_{\mathbf k}.
  \label{eq:curv-dphi-projection}
\end{align}
The zero mode is excluded from this projection and is set to
\begin{equation}
  \Phi_{\mathbf 0}
  =
  \Psi_{\mathbf 0}
  =
  \Psi'_{\mathbf 0}
  =
  0 .
  \label{eq:curv-zero-mode}
\end{equation}

Radiation perturbations are not determined by the constraint projection.
Instead, they are evolved as linear matter degrees of freedom. Neglecting
radiation shear and collision terms, their evolution equations are
\begin{align}
  \delta'_{r,\mathbf k}
  &=
  -\frac43\theta_{r,\mathbf k}
  +
  4\Psi'_{\mathbf k},
  \label{eq:curv-delta-r-eq} \\
  \theta'_{r,\mathbf k}
  &=
  k^2
  \left(
    \frac14\delta_{r,\mathbf k}
    +
    \Phi_{\mathbf k}
  \right).
  \label{eq:curv-theta-r-eq}
\end{align}
Time evolution is performed using a {second-order Runge--Kutta (RK) scheme}. At time
step \(n\), the radiation variables are used to project
\(\Phi_{\mathbf k}^{(1)}\), \(\Psi_{\mathbf k}^{(1)}\), and
\(\Psi_{\mathbf k}^{\prime(1)}\). The superscript \((1)\) labels this first
{RK stage}, and the corresponding right-hand sides, or slopes, of the
\(\delta_r\) and \(\theta_r\) equations are denoted by \(D_\delta^{(1)}\)
and \(D_\theta^{(1)}\):
\begin{align}
  D_\delta^{(1)}
  &=
  -\frac43\theta_{r,\mathbf k}
  +
  4\Psi_{\mathbf k}^{\prime(1)},
  \notag \\
  D_\theta^{(1)}
  &=
  k^2
  \left(
    \frac14\delta_{r,\mathbf k}
    +
    \Phi_{\mathbf k}^{(1)}
  \right).
  \label{eq:curv-rk2-slope1}
\end{align}
Using a conformal-time step \(\Delta\tau\), the first-stage slopes yield the
forward-Euler predictions, denoted by the label \({\rm pred}\),
\begin{align}
  \delta_{r,\mathbf k}^{\rm pred}
  &=
  \delta_{r,\mathbf k}
  +
  \Delta\tau D_\delta^{(1)},
  \notag \\
  \theta_{r,\mathbf k}^{\rm pred}
  &=
  \theta_{r,\mathbf k}
  +
  \Delta\tau D_\theta^{(1)} .
  \label{eq:curv-rk2-predictor}
\end{align}
The metric is then projected again using the predicted radiation sources. This
defines the second {RK stage} and the corresponding slopes
\(D_\delta^{(2)}\) and \(D_\theta^{(2)}\):
\begin{align}
  D_\delta^{(2)}
  &=
  -\frac43\theta_{r,\mathbf k}^{\rm pred}
  +
  4\Psi_{\mathbf k}^{\prime\,\rm pred},
  \notag \\
  D_\theta^{(2)}
  &=
  k^2
  \left(
    \frac14\delta_{r,\mathbf k}^{\rm pred}
    +
    \Phi_{\mathbf k}^{\rm pred}
  \right).
  \label{eq:curv-rk2-slope2}
\end{align}
Combining the two stages gives the radiation variables at the next time step,
denoted by \(n+1\):
\begin{align}
  \delta_{r,\mathbf k}^{\,n+1}
  &=
  \delta_{r,\mathbf k}^{\,n}
  +
  \frac{\Delta\tau}{2}
  \left(
    D_\delta^{(1)}
    +
    D_\delta^{(2)}
  \right),
  \label{eq:curv-rk2-delta-final} \\
  \theta_{r,\mathbf k}^{\,n+1}
  &=
  \theta_{r,\mathbf k}^{\,n}
  +
  \frac{\Delta\tau}{2}
  \left(
    D_\theta^{(1)}
    +
    D_\theta^{(2)}
  \right).
  \label{eq:curv-rk2-theta-final}
\end{align}
A final constraint projection using the updated radiation perturbations yields
\(\Phi_{\mathbf k}^{\,n+1}\), \(\Psi_{\mathbf k}^{\,n+1}\), and
\(\Psi_{\mathbf k}^{\prime\,n+1}\) at the end of the time step.

\subsection{Constraint reconstruction at bubble nucleation}
\label{app:curv-nucleation}

Bubble nucleation changes the scalar-field configuration instantaneously and
therefore modifies \(\delta\rho_\phi\), \(\delta p_\phi\), and
\(\Pi^S_\phi\). To match the metric perturbations to the matter sources before
and after nucleation, the module stores the scalar-field configuration before
the bubble is inserted. The differences across nucleation are then
\begin{equation}
  \begin{gathered}
  \Delta\delta\rho_{\phi,\mathbf k},\quad
  \Delta\Pi^S_{\phi,\mathbf k}.
  \end{gathered}
  \label{eq:curv-bubble-delta-source}
\end{equation}
These increments are used primarily to monitor the local source change induced
by bubble insertion. The metric used in the subsequent evolution is not,
however, determined by these increments alone. Instead, after nucleation the
full scalar-field sources and the pre-existing radiation perturbations are
recomputed, and Eqs.~\eqref{eq:curv-phi-projection}--\eqref{eq:curv-dphi-projection}
are applied again to obtain the new \(\Phi\), \(\Psi\), and \(\Psi'\).

Within the static bubble-insertion approximation adopted here, a bubble is
inserted without an accompanying perturbation to either the scalar-field
momentum or the radiation velocity. Accordingly, the nucleation routine does
not introduce independent increments \(\Delta q_\phi\) or
\(\Delta\theta_r\). The post-nucleation value of \(\Psi'_{\mathbf k}\) is
then obtained solely by reapplying the constraint projection to the
post-nucleation sources.

\section{Gravitational-wave computation}
\label{app:gravitational-wave-module}

This section presents the prescription used to evolve and measure tensor
perturbations in our simulations.  Within a single simulation, we
distinguish {GWs} generated directly by the phase transition,
{GWs} induced by first-order scalar metric perturbations, and the
total {GW} spectrum obtained by adding the two coherently.  We
evolve the six independent components of a symmetric
tensor,
\begin{equation}
  u_{ij}
  =
  u_{ji},
  \qquad
  ij
  =
  11,12,13,22,23,33 ,
  \label{eq:gw-six-components}
\end{equation}
and apply a {TT projection} to the Fourier modes when measuring
the spectra.  Here \(u_{ij}\) denotes the auxiliary tensor field evolved
numerically.  For every channel \(X\), the physical TT perturbation is defined
uniformly by
\begin{equation}
  h_{ij}^{X}
  =g\Lambda_{ij}^{\ lm}u_{lm}^{X},
  \qquad
  g\equiv
  \left(\frac{f_*}{M_{\rm Pl}}\right)^2,
  \label{eq:gw-auxiliary-normalization}
\end{equation}
where \(\Lambda_{ij}^{\ lm}\) is the TT projector.  This relation is used for both
the direct and induced channels, allowing their auxiliary fields to be combined at
the field level with a common normalization.

We evolve three tensor channels separately,
\begin{equation}
  h_{ij}^{\mathrm{PT}},
  \qquad
  h_{ij}^{\mathrm{ind,geo}},
  \qquad
  h_{ij}^{\mathrm{ind,geo+rad}}.
  \label{eq:gw-three-channels}
\end{equation}
Here \(h_{ij}^{\mathrm{PT}}\) is the phase-transition {GW} driven
directly by the anisotropic stress of the lattice matter fields;
\(h_{ij}^{\mathrm{ind,geo}}\) is the {IGW} driven only by the
second-order geometric source built from the first-order scalar metric
perturbations; and \(h_{ij}^{\mathrm{ind,geo+rad}}\) supplements the geometric
source with a radiation-fluid velocity source.  All three tensor fields start
from zero initial conditions.

\subsection{Tensor evolution and source decomposition}
\label{app:gw-parameters-fields}

We evolve three tensor responses separately: the phase-transition,
purely geometric induced, and geometric-plus-radiation-velocity induced channels.
They can be used separately for diagnostic comparisons or combined to construct
the total spectrum.  The total spectrum is not obtained from a mixed source but by
adding the corresponding tensor fields at the field level when it is measured.
Because the tensor equation is linear in \(h_{ij}\), we form the total field at
measurement time by adding the separately evolved responses; squaring this sum
retains coherent cross terms between their sources.

The induced channels require the first-order scalar metric perturbations \(\Phi\)
and \(\Psi\), together with the radiation-velocity perturbation.  They therefore
are physically meaningful only when the metric--radiation sector is evolved
simultaneously.  All tensor channels are
evolved on a three-dimensional lattice.  One or more channels may be evolved
simultaneously without changing the primary scalar-field dynamics or the
definitions of the individual channels given below.

We use the dimensionless program variables and source contributions introduced
in Sec.~\ref{sec:numerical-setup} and the same tensor normalization as in
Eq.~\eqref{eq:setup-metric-tensor}.
Throughout this subsection, hats on
coordinates, derivatives, \(\mathcal H\), \(v_r\), and
\(S_{ij}^{\rm s,g,r}\) are suppressed; hats on the radiation density and
pressure are retained. The sources are the dimensionless forms
\(\widehat S_{ij}^{X}=S_{ij}^{X}/\omega_*^2\) of the physical contributions
in Sec.~\ref{sec:setup-tensor}, not a separate source convention.
The phase-transition channel obeys
\begin{equation}
  h_{ij}^{\rm PT\,\prime\prime}
  +2\mathcal H h_{ij}^{\rm PT\,\prime}
  -\nabla^2 h_{ij}^{\rm PT}
  =\Lambda_{ij}^{lm}S_{lm}^{\rm s},
  \label{eq:gw-pt-continuum}
\end{equation}
where the scalar-field source is
\begin{equation}
  S_{ij}^{\rm s}=2g\,\partial_i\varphi\,\partial_j\varphi,
  \label{eq:gw-scalar-pi}
\end{equation}
in continuum notation. Its lattice implementation uses forward differences.
The homogeneous radiation background provides no local TT stress to this
direct channel; the radiation-velocity contribution is treated separately below.
We denote the auxiliary-tensor rescaling exponent by \(\alpha_u\).
The discrete evolution of the six symmetric components uses
\begin{equation}
  K_{ij}^{\rm PT}
  =a^{1+\alpha_u}\left[
    \nabla^2u_{ij}^{\rm PT}+\frac{1}{g}S_{ij}^{\rm s}
  \right],
  \label{eq:gw-pt-kernel}
\end{equation}
as the right-hand side of the tensor-momentum equation.
The exponent \(\alpha_u\) must not be confused with the phase-transition
strength \(\alpha\).

The scalar metric potentials in the induced channels come from the scalar
metric--radiation perturbation prescription and use the conformal Newtonian
gauge of Eq.~\eqref{eq:curv-newtonian-gauge}.
The geometric source retains the second-order geometric nonlinearities of
the first-order scalar metric perturbations. Defining
\(\partial_{ij}\equiv\partial_i\partial_j\), it is \cite{Baumann:2007zm}
\begin{equation}
  \begin{aligned}
  S_{ij}^{\rm g}=-2\big[&
    2\Phi\,\partial_{ij}\Phi
    -2\Psi\,\partial_{ij}\Phi
    +4\Psi\,\partial_{ij}\Psi
    +(\partial_i\Phi)(\partial_j\Phi)\\
    &-(\partial_i\Phi)(\partial_j\Psi)
    -(\partial_i\Psi)(\partial_j\Phi)
    +3(\partial_i\Psi)(\partial_j\Psi)\big].
  \end{aligned}
  \label{eq:gw-geo-source}
\end{equation}
The corresponding tensor equation is
\begin{equation}
  h_{ij}^{\mathrm{ind,geo}\,\prime\prime}
  +2\mathcal H h_{ij}^{\mathrm{ind,geo}\,\prime}
  -\nabla^2h_{ij}^{\mathrm{ind,geo}}
  =\Lambda_{ij}^{lm}S_{lm}^{\rm g}.
  \label{eq:gw-geo-equation}
\end{equation}
We construct this source in real space for the six symmetric tensor components.
The discrete evolution term is
\begin{equation}
  K_{ij}^{\mathrm{ind,geo}}
  =a^{1+\alpha_u}\left[
    \nabla^2u_{ij}^{\mathrm{ind,geo}}+\frac{1}{g}S_{ij}^{\rm g}
  \right].
  \label{eq:gw-geo-kernel}
\end{equation}
Here \(K_{ij}\) is the discrete right-hand side of the tensor-momentum equation.
The gradients and Hessian in the geometric source use centered differences,
tensor propagation uses the nearest-neighbor lattice Laplacian, and the Fourier
constraint projection uses the continuum mode label \(k_i=2\pi n_i/L\).
These operators agree in the infrared but differ in the ultraviolet.

The geometric-plus-radiation-velocity channel adds \(S_{ij}^{\rm r}\)
to \(S_{ij}^{\rm g}\). The radiation-velocity divergence evolved by the
scalar metric--radiation perturbation prescription is denoted by \(\theta_r\).
For a scalar velocity mode, we use
\begin{equation}
  v_{r,i}=-\partial_i v_r,\qquad
  \theta_r=\partial_i v_r^i=-\nabla^2v_r,
  \label{eq:gw-scalar-velocity-convention}
\end{equation}
so that the velocity potential in Fourier space is
\begin{equation}
  v_{r,\mathbf k}=\frac{\theta_{r,\mathbf k}}{k^2},\qquad k\neq0,
  \label{eq:gw-vr-definition}
\end{equation}
with zero mode
\begin{equation}
  v_{r,\mathbf 0}=0.
  \label{eq:gw-vr-zero-mode}
\end{equation}
Writing the background density and pressure in lattice units as
\(\widehat\rho_r=\bar\rho_r^{\rm phys}/(f_*^2\omega_*^2)\) and
\(\widehat P_r=\bar P_r^{\rm phys}/(f_*^2\omega_*^2)\), the radiation source is
\begin{equation}
  S_{ij}^{\rm r}
  =2g a^2(\widehat\rho_r+\widehat P_r)
  (\partial_i v_r)(\partial_j v_r).
  \label{eq:gw-rad-source-general}
\end{equation}
For radiation, \(\widehat P_r=\widehat\rho_r/3\), giving
\begin{equation}
  S_{ij}^{\rm r}
  =\frac{8}{3}g a^2\widehat\rho_r
  (\partial_i v_r)(\partial_j v_r).
  \label{eq:gw-rad-source-radiation}
\end{equation}
The constants are defined by
\begin{equation}
  \kappa^2=8\pi G=M_{\rm Pl}^{-2},\qquad
  g=\kappa^2 f_*^2=\left(\frac{f_*}{M_{\rm Pl}}\right)^2.
  \label{eq:gw-kappa}
\end{equation}
Thus \(\kappa^2\bar\rho_r^{\rm phys}/\omega_*^2=g\widehat\rho_r\),
consistent with the normalization in Eq.~\eqref{eq:setup-dimensionless-potential}.
The combined induced channel obeys
\begin{equation}
  \begin{aligned}
  h_{ij}^{\mathrm{ind,geo+rad}\,\prime\prime}
  &+2\mathcal H h_{ij}^{\mathrm{ind,geo+rad}\,\prime}
  -\nabla^2h_{ij}^{\mathrm{ind,geo+rad}}\\
  &=\Lambda_{ij}^{lm}\left(S_{lm}^{\rm g}+S_{lm}^{\rm r}\right),
  \end{aligned}
  \label{eq:gw-geo-rad-equation}
\end{equation}
and its discrete evolution term is
\begin{equation}
  K_{ij}^{\mathrm{ind,geo+rad}}
  =a^{1+\alpha_u}\left[
    \nabla^2u_{ij}^{\mathrm{ind,geo+rad}}
    +\frac{1}{g}\left(S_{ij}^{\rm g}+S_{ij}^{\rm r}\right)
  \right].
  \label{eq:gw-geo-rad-kernel}
\end{equation}
All three sources include their complete prefactors in the physical tensor
equation. The common factor \(1/g\) in the auxiliary-tensor equations follows
from \(h_{ij}^{X}=g\Lambda_{ij}^{lm}u_{lm}^{X}\).

This channel does not include the second-order
scalar-field velocity or pressure sources.  The channel separation is used to
compare explicitly the relative roles of the geometric, radiation-velocity, and
direct matter-stress terms, but it does not imply that all sources in the complete
second-order Einstein--matter system have been included.

\label{app:gw-time-evolution}

Each channel has its own tensor field \(u_{ij}^{X}\) and conjugate variable
\(\pi_{ij}^{X}\), where
\begin{equation}
  X
  \in
  \{
    \mathrm{PT},
    \mathrm{ind,geo},
    \mathrm{ind,geo+rad}
  \}.
  \label{eq:gw-channel-index}
\end{equation}
With the rescaled conjugate variable \(\pi_{ij}^{X}\), the field and momentum satisfy
\begin{equation}
  u_{ij}^{X\,\prime}
  =
  a^{\alpha_u-3}\pi_{ij}^{X},
  \label{eq:gw-field-update}
\end{equation}
and
\begin{equation}
  \pi_{ij}^{X\,\prime}
  =
  K_{ij}^{X}.
  \label{eq:gw-momentum-update}
\end{equation}
All three channels are integrated with the same staggered symplectic scheme and
differ only through the sources in
Eqs.~\eqref{eq:gw-pt-kernel}, \eqref{eq:gw-geo-kernel}, and
\eqref{eq:gw-geo-rad-kernel}.

{We update \(S_{ij}^{\rm g}\) and \(S_{ij}^{\rm r}\) after each}
perturbation time step.  The induced channels therefore use the current
constraint-projected values of \(\Phi\), \(\Psi\), and the radiation-velocity
potential, rather than independently evolved scalar potentials.

Because the {GW} equation is linear in \(h_{ij}\), tensor fields and
their conjugate momenta generated by different sources can be added at the field
level.  The {GW} energy spectrum is measured from the TT-projected
tensor momentum.  For any two channels \(A\) and \(B\), we therefore first
construct
\begin{equation}
  \pi_{ij}^{\mathrm{total,TT}}
  =
  \pi_{ij}^{A,\mathrm{TT}}
  +
  \pi_{ij}^{B,\mathrm{TT}}.
  \label{eq:gw-cross-term}
\end{equation}
The corresponding shell average satisfies
\begin{equation}
  \begin{aligned}
  \left\langle |\pi^{\mathrm{total,TT}}|^2\right\rangle_b
  &=
  \left\langle |\pi^{A,\mathrm{TT}}|^2\right\rangle_b
  +
  \left\langle |\pi^{B,\mathrm{TT}}|^2\right\rangle_b
  +
  2\,\mathrm{Re}\left\langle
  \pi_{ij}^{A,\mathrm{TT}}
  \pi_{ij}^{B,\mathrm{TT}\,*}
  \right\rangle_b .
  \end{aligned}
  \label{eq:gw-momentum-cross-term}
\end{equation}
The total energy spectrum consequently includes the coherent cross term
automatically.  The numerical scheme first evolves the channels independently and
then performs a field-level sum of both the tensor fields and tensor momenta before
measuring the total spectrum.

When the geometric-plus-radiation-velocity induced channel is included, the total
tensor field is defined as
\begin{equation}
  h_{ij}^{\mathrm{total}}
  =
  h_{ij}^{\mathrm{PT}}
  +
  h_{ij}^{\mathrm{ind,geo+rad}}.
  \label{eq:gw-total-fluid}
\end{equation}
The purely geometric channel \(h_{ij}^{\mathrm{ind,geo}}\) is then retained only
for comparison and is not added separately to \(h_{ij}^{\mathrm{total}}\),
thereby avoiding double counting of the geometric source.  If only the purely
geometric induced channel is included, then
\begin{equation}
  h_{ij}^{\mathrm{total}}
  =
  h_{ij}^{\mathrm{PT}}
  +
  h_{ij}^{\mathrm{ind,geo}}.
  \label{eq:gw-total-geo}
\end{equation}
If the phase-transition channel is absent, the total spectrum reduces to that of
the retained induced channel.

\subsection{Spectral measurement}
\label{app:gw-spectrum-output}

When measuring the spectrum, a TT projection is applied in Fourier space to the
tensor momentum \(\pi_{ij}^{X}\) of every channel.  The projector is constructed
from the lattice momentum \(k_i^{\mathrm{lat}}\),
\begin{equation}
  P_{ij}
  =
  \delta_{ij}
  -
  \frac{
    k_i^{\mathrm{lat}\,\ast}k_j^{\mathrm{lat}}
  }{
    |{\bf k}^{\mathrm{lat}}|^2
  },
  \label{eq:gw-transverse-projector}
\end{equation}
and then combined into the tensor TT projector.  Different definitions of the
lattice momentum correspond to different discrete projection prescriptions.

The numerical measurement first performs a radial shell average of the quadratic
form of the TT-projected tensor momentum.  To make the rescaling of the numerical
variables explicit, define the projected quantity in wavenumber shell \(b\) by
\begin{equation}
  \mathcal Q_b^X
  \equiv
  \left\langle
    a^{2(\alpha_u-3)}
    \pi_{ij}^{X,\mathrm{TT}}({\bf k})
    \pi_{ij}^{X,\mathrm{TT}}({\bf k})^\ast
  \right\rangle_b ,
  \label{eq:gw-shell-average}
\end{equation}
where repeated spatial indices are summed and \(\langle\cdots\rangle_b\) denotes
a shell average.  The resulting dimensionless {GW} energy spectrum
is
\begin{equation}
  \Omega_{\rm GW}^{X}(k_b)
  =
  \frac{(k_b\Delta x)^3}
  {8\pi^2N^3a^{2\alpha_u}\rho_{\rm tot}}
  \left(
    \frac{f_*}{M_{\rm Pl}}
  \right)^2
  \mathcal Q_b^X ,
  \label{eq:gw-spectrum-normalization}
\end{equation}
where \(N\) is the number of lattice sites per side, \(\Delta x\) is the comoving
lattice spacing, \(k_b\) is the central comoving wavenumber of the shell, and
\begin{equation}
  \rho_{\rm tot}=\bar\rho_\phi+\bar\rho_r,
  \label{eq:gw-total-background-density}
\end{equation}
explicitly includes the homogeneous radiation background.  For the total
channel, \(\mathcal Q_b^{\rm total}\) is calculated from the summed tensor momentum
in Eq.~\eqref{eq:gw-cross-term}.  The cross term in
Eq.~\eqref{eq:gw-momentum-cross-term} therefore enters
\(\Omega_{\rm GW}^{\rm total}\) with the same normalization as the autocorrelation
terms.

For every channel, the integrated energy density is obtained from
\(\rho_{\rm GW}=\Omega_{\rm GW}\rho_{\rm tot}\), using the same total background
energy density, including radiation, as in the spectral estimator.

The physical source decomposition adopted here is
\begin{equation}
  \begin{aligned}
  h_{ij}^{\mathrm{PT}}
  &\leftarrow
    {S_{ij}^{\rm s}},
  \\
  h_{ij}^{\mathrm{ind,geo}}
  &\leftarrow
  {S_{ij}^{\rm g}},
  \\
  h_{ij}^{\mathrm{ind,geo+rad}}
  &\leftarrow
  {S_{ij}^{\rm g}}
  +
  {S_{ij}^{\rm r}}.
  \end{aligned}
  \label{eq:gw-channel-summary}
\end{equation}
This division operationally separates the direct matter-stress channel from the
geometric-plus-radiation-velocity channel used here, facilitating comparison of
their respective roles.  Their combination remains subject to the common
auxiliary-field normalization and second-order counting convention described
above.  The fluid term in the geometric-plus-radiation-velocity channel contains
only the square of gradients of the radiation-velocity potential.  It does not
contain a scalar-field velocity source, a pressure term, or a matter term
constructed from the total momentum variable \(U=\Psi'+\mathcal H\Phi\).

\section{Power spectra and probability distributions of scalar perturbations}
\label{app:power-spectrum-pdf-module}

Perturbations generated during the phase transition have a nontrivial scale
dependence and may also exhibit non-Gaussian statistics.  Power spectra and
one-point {PDFs} provide complementary statistical
information.  The former describe how the variance of the fluctuations is
distributed among Fourier scales and can therefore identify macroscopic
characteristic scales such as the mean bubble separation and the comoving Hubble
radius.  A {PDF}, by contrast, retains higher-order
one-point information that is not fixed by the power spectrum and can be used
to test whether the density perturbations generated by the transition are skewed
or heavy-tailed.

This section uses the conformal Newtonian gauge of
Eq.~\eqref{eq:curv-newtonian-gauge}, where \(\Phi\) is the lapse (Newtonian)
potential, \(\Psi\) is the spatial-curvature potential, and
\(\mathcal H=a'/a\) is the conformal Hubble rate. We describe how the
evolved fields are used to construct the metric power spectra and the smoothed
comoving-density PDF, and discuss the physical meaning of these complementary
statistics.  The power spectra of \(\Phi\) and \(\Psi\) are the primary scalar
two-point observables, the slip spectrum isolates the anisotropic-stress
response, and the radiation fields connect the spatial morphology to the
metric evolution.  The comoving-density PDF and its standard deviation and
skewness provide the main one-point non-Gaussian observables, while \(\zeta\)
is retained as a derived diagnostic.

\subsection{Power spectrum and its physical interpretation}
\label{app:pspdf-power-spectrum}

For a statistically homogeneous and isotropic real scalar field \(X({\bf x})\)
whose spatial mean has been subtracted, the continuum Fourier-space power spectrum
is defined by
\begin{equation}
  \left\langle X({\bf k})X^*({\bf k}')\right\rangle
  =(2\pi)^3\delta^{(3)}({\bf k}-{\bf k}')P_X(k).
  \label{eq:pspdf-continuum-power}
\end{equation}
We primarily use the dimensionless power spectrum
\begin{equation}
  \Delta_X^2(k)=\frac{k^3}{2\pi^2}P_X(k),
  \label{eq:pspdf-dimensionless-power}
\end{equation}
which gives the contribution to the real-space variance per logarithmic interval
in wavenumber:
\begin{equation}
  \left\langle (X-\langle X\rangle)^2\right\rangle
  =\int d\ln k\,\Delta_X^2(k).
  \label{eq:pspdf-variance-from-power}
\end{equation}
The peak of \(\Delta_X^2(k)\) therefore marks the scale contributing most strongly
to the variance of that fluctuation.  It contains the same two-point information
as the real-space two-point correlation function.

The simulation uses periodic comoving coordinates in a box of side
\(L\), with \(N\) lattice points in each direction.  The lowest nonzero
comoving wavenumber and the lattice spacing are therefore
{
\begin{equation}
  k_{\rm IR}=\frac{2\pi}{L},
  \qquad
  \Delta x=\frac{L}{N}.
  \label{eq:pspdf-box-scales}
\end{equation}
}
The horizontal axis of every spectrum is the comoving wavenumber \(k\); the
corresponding physical wavenumber is
\begin{equation}
  k_{\rm phys}=\frac{k}{a}.
  \label{eq:pspdf-physical-k}
\end{equation}
In a finite box, the spatial average is used as an estimator of the ensemble
average, and discrete Fourier modes with similar \(|{\bf k}|\) are averaged over
spherical shells.  The discrete
Fourier convention is
\begin{equation}
  \widetilde X_{\bf n}
  =\sum_{\bf j}X({\bf x}_{\bf j})
  e^{-i{\bf k}_{\bf n}\cdot{\bf x}_{\bf j}},
  \label{eq:pspdf-discrete-fourier}
\end{equation}
with no division by the total number of lattice points in the transform.  The
dimensionless spectral estimator used here is
\begin{equation}
  \widehat{\Delta_X^2}(k_b)
  =\frac{(k_b\Delta x)^3}{2\pi^2N^3}
  \frac{1}{N_{\rm mode}(k_b)}
  \sum_{{\bf n}\in\mathcal S_b}
  |\widetilde X_{\bf n}|^2,
  \label{eq:pspdf-lattice-estimator}
\end{equation}
where \(\mathcal S_b\) is the set of modes in radial shell \(b\), \(N_{\rm mode}(k_b)\) is the number
of Fourier modes in that shell, and \(k_b\) is its representative wavenumber.
Equation~\eqref{eq:pspdf-lattice-estimator} is the finite-periodic-box
discretization of the continuum definition in
Eq.~\eqref{eq:pspdf-dimensionless-power} and uses the \textsc{CosmoLattice} power-spectrum
normalization convention \cite{Figueroa:2020rrl,Figueroa:2021yhd}.

\subsection{Scalar statistical variables}
\label{app:pspdf-measured-fields}

The following five scalar quantities are related, although they play different
roles in the analysis presented here:
\begin{equation}
  \Phi,
  \qquad \Psi,
  \qquad \Phi-\Psi,
  \qquad \zeta,
  \qquad \delta_c.
  \label{eq:pspdf-measured-list}
\end{equation}
The potentials \(\Phi\) and \(\Psi\) directly describe the scalar gravitational
field in Newtonian gauge.  Their difference \(\Phi-\Psi\) is the gravitational
slip, which in our system reflects the anisotropic stress generated by gradients
of the phase-transition scalar field.  Comparing these three spectra separates
the total scalar gravitational potentials from the component controlled by
anisotropic stress.

We therefore report \(\Delta_\Phi^2(k)\) and
\(\Delta_\Psi^2(k)\) first and use \(\Delta_{\Phi-\Psi}^2(k)\) to isolate the
anisotropic-stress response.  The evolved \(\delta_r\) and \(\theta_r\) fields
provide the accompanying radiation morphology and enter the tensor source.
The variable \(\zeta\) remains a derived diagnostic.  The comoving density
contrast \(\delta_c\) does not replace the potential spectra; its smoothed
one-point PDF instead supplies the complementary non-Gaussian density
observable studied in Sec.~\ref{sec:results-density-pdf}.

The curvature perturbation \(\zeta\) on uniform-total-density hypersurfaces
describes the spatial curvature.  The total density perturbation and total
background enthalpy density are
\begin{align}
  \delta\rho_{\rm tot}
  &=\delta\rho_\phi+\bar\rho_r\delta_r,
  \label{eq:pspdf-total-density-perturbation}\\
  \bar\rho_{\rm tot}+\bar p_{\rm tot}
  &=\bar\rho_\phi+\bar p_\phi+\frac{4}{3}\bar\rho_r.
  \label{eq:pspdf-total-enthalpy}
\end{align}
Using the background energy-conservation equation
\begin{equation}
  \bar\rho'_{\rm tot}
  =-3\mathcal H(\bar\rho_{\rm tot}+\bar p_{\rm tot}),
  \label{eq:pspdf-background-conservation}
\end{equation}
it can be written in Newtonian gauge as
\begin{equation}
  \zeta
  =-\Psi
  +\frac{\delta\rho_\phi+\bar\rho_r\delta_r}
  {3\left(\bar\rho_\phi+\bar p_\phi+\frac43\bar\rho_r\right)}.
  \label{eq:pspdf-zeta}
\end{equation}
The quantity \(\zeta\) provides a natural connection between phase-transition
dynamics and the usual language of cosmological initial perturbations.  Local
energy conservation gives its first-order evolution equation,
\begin{equation}
  \zeta'
  =-\frac{\mathcal H}{\bar\rho_{\rm tot}+\bar p_{\rm tot}}
  \delta p_{\rm nad}
  -\frac13\nabla^2 V_{\rm ud},
  \label{eq:pspdf-zeta-evolution}
\end{equation}
where \(\delta p_{\rm nad}\) is the total nonadiabatic pressure perturbation and
\(V_{\rm ud}\) denotes a velocity or shear combination on uniform-density
hypersurfaces.  Thus, for \(k\ll\mathcal H\) and negligible
\(\delta p_{\rm nad}\), \(\zeta\) tends to be conserved
\cite{Wands:2000dp}.  Anisotropic stress controls \(\Phi-\Psi\) and can affect
\(\zeta\) indirectly through the metric and finite-wavenumber dynamics, but it
does not constitute an independent superhorizon source in
Eq.~\eqref{eq:pspdf-zeta-evolution}.  We therefore use the stability of the
superhorizon part of \(\Delta_\zeta^2(k)\) after the phase transition only as a
derived diagnostic; subhorizon modes can continue to evolve through the gradient
term.  It is also important that a discrete nucleation event changes the scalar
field instantaneously without explicitly inserting a smooth energy-
transfer term.  At that instant, the smooth background conservation relation used
to derive Eq.~\eqref{eq:pspdf-zeta} does not hold exactly.  Near an event,
\(\zeta\) should therefore be regarded as an algebraically constructed diagnostic
within the injection prescription, rather than as a rigorously conserved
gauge-invariant quantity.

For density statistics, we use the total density contrast in comoving gauge.  The
sign and scale-factor convention for the momentum potential is
\begin{equation}
  \delta T^0{}_{i,{\bf k}}
  =-\frac{ik_i}{a^2}q_{\bf k},
  \qquad
  q_{{\bf k}}
  =q_{\phi,{\bf k}}
  +\frac{4a^2\bar\rho_r}{3k^2}\theta_{r,{\bf k}}.
  \label{eq:pspdf-total-momentum-potential}
\end{equation}
For nonzero Fourier modes, we use
\begin{equation}
  \begin{aligned}
  \delta_{c,{\bf k}}
  =\frac{1}{\bar\rho_\phi+\bar\rho_r}
  \Bigg\{&\delta\rho_{\phi,{\bf k}}
  +\bar\rho_r\delta_{r,{\bf k}}+\frac{3\mathcal H}{a^2}\left[
  q_{\phi,{\bf k}}
  +\frac{4a^2\bar\rho_r}{3k^2}\theta_{r,{\bf k}}
  \right]\Bigg\}.
  \end{aligned}
  \label{eq:pspdf-delta-comoving}
\end{equation}
Here \(q_{\phi,{\bf k}}\) is the longitudinal scalar-field momentum potential and
\(\theta_{r,{\bf k}}\) is the radiation-velocity divergence.  The spatially
averaged density is already included in the background, so the zero mode is
defined by
\begin{equation}
  \delta_{c,{\bf 0}}=0.
  \label{eq:pspdf-delta-comoving-zero}
\end{equation}
The quantity \(\delta_c\) is defined on a slicing comoving with the total matter
flow, on which the total momentum potential vanishes.  The momentum term in
Eq.~\eqref{eq:pspdf-delta-comoving} implements
the first-order gauge transformation from Newtonian slicing to comoving slicing.
The resulting variable is more directly connected to the density measured by a
comoving observer and to conventional horizon-scale density statistics.

Equations~\eqref{eq:pspdf-zeta} and \eqref{eq:pspdf-delta-comoving} use first-order
gauge transformations.  They describe statistics produced by a nonlinear phase-
transition background within a framework of linear scalar metric perturbations;
they are not rigorously nonlinear variables under a complete second-order gauge
transformation.  The PDF derived from \(\delta_c\) can characterize the
non-Gaussian shape generated by the phase
transition.  We do not, however, use it here to calculate a nonlinear collapse
threshold or {PBH abundance}.  If the PDF tail reaches
\(|\delta_c|\sim\mathcal O(1)\), the validity of the first-order gauge
transformation and linear metric treatment must be reassessed.

\subsection{Smoothed density field and its probability distribution}
\label{app:pspdf-smoothed-pdf}

A density PDF must be defined together with a smoothing scale.  An unsmoothed
lattice density is highly sensitive to the bubble-wall thickness, ultraviolet
gradients, and lattice resolution, and is difficult to associate with a well-
defined cosmological scale.  For completeness, we use a Gaussian window,
\begin{equation}
  W(kR)=\exp\!\left(-\frac12k^2R^2\right),
  \label{eq:pspdf-gaussian-window}
\end{equation}
and define
\begin{equation}
  \delta_c^{(R)}({\bf x})
  =\mathcal F^{-1}\!\left[
  W(kR)\delta_{c,{\bf k}}
  \right].
  \label{eq:pspdf-smoothed-delta}
\end{equation}
Because \(k\) is a comoving wavenumber, \(R\) is also a comoving scale.  The
variance of the smoothed field should satisfy
\begin{equation}
  \sigma_R^2
  =\int d\ln k\,
  \Delta_{\delta_c}^2(k)W^2(kR),
  \label{eq:pspdf-smoothed-variance}
\end{equation}
which provides a direct consistency check between the power spectrum and the
real-space PDF.

To compare with previous studies, we evaluate the PDF
\(p[\delta_c^{(\mathcal H_{\max}^{-1})}]\), where
\begin{equation}
  \mathcal H_{\max}
  \equiv \max_{\tau'\leq\tau}\mathcal H(\tau').
  \label{eq:pspdf-horizon-scale}
\end{equation}
Here the argument of \(\delta_c\) denotes the comoving smoothing radius.
For the \(\alpha\geq1\) cases considered here, the background undergoes a
brief inflationary stage followed by radiation domination. The comoving
Hubble radius therefore decreases and subsequently grows, making its
minimum value, \(\mathcal H_{\max}^{-1}\), a suitable scale for studying
the density perturbations just after the transition.

Our parameter scan also includes \(\alpha<1\) cases, for which the
background remains radiation dominated and the comoving Hubble radius
grows monotonically. In these cases, \(\mathcal H_{\max}^{-1}\) corresponds
to an early-time scale rather than the horizon scale at transition
completion. To compare the PDFs across all parameter choices, we
therefore smooth the density field on the comoving Hubble radius
\(\mathcal H^{-1}(\tau_{0.99})\) at transition completion. We define
\(\tau_{0.99}\) as the first simulation time at which
\begin{equation}
  x_\phi(\tau)\equiv\langle\phi(\tau,{\bf x})\rangle\geq0.99,
  \label{eq:pspdf-xphi-completion}
\end{equation}
where the brackets denote a volume average and the field values in the
false and true vacua are normalized to zero and unity, respectively.
We thus compare \(p[\delta_c^{(\mathcal H^{-1}(\tau_{0.99}))}]\) at the same
transition stage, using the corresponding horizon scale for each
parameter set.

A normalized histogram is constructed from the spatial distribution of the
smoothed field.  If bin \(b\) has width \(\Delta\delta\), contains \(C_b\) samples,
and the total number of samples is \(N_{\rm s}\), then
\begin{equation}
  \widehat p_b
  =\frac{C_b}{N_{\rm s}\Delta\delta},
  \qquad
  \sum_b\widehat p_b\Delta\delta=1.
  \label{eq:pspdf-histogram-estimator}
\end{equation}
{In addition to the full PDF, we characterize its width by the
standard deviation of the smoothed comoving-density contrast,}
\begin{equation}
  {\sigma_R
  \equiv\left\langle
  \left[\delta_c^{(R)}-\langle\delta_c^{(R)}\rangle\right]^2
  \right\rangle^{1/2}.}
  \label{eq:pspdf-sigma}
\end{equation}
{Here the brackets denote the spatial average over the lattice at
the measurement time.  Thus \(\sigma_R\) is the root-mean-square spread of the
Gaussian-smoothed field about its spatial mean.  At fixed smoothing radius, a
larger \(\sigma_R\) corresponds to a broader one-point distribution and hence to
a larger typical amplitude of the density fluctuations.  The full PDF is kept
alongside \(\sigma_R\) to retain information about its detailed shape and tails.}

\ifdefined\INTEGRATEDMANUSCRIPT
\else
\renewcommand{\refname}{References}
\bibliographystyle{unsrt}
\bibliography{paper_references}

\end{document}
\fi

\bibliographystyle{JHEP}
\bibliography{paper_references}
\end{document}